\documentclass[fleqn,usenatbib]{mnras}

\usepackage{float}

\usepackage{newtxtext,newtxmath}
\usepackage{xcolor}
\usepackage{anyfontsize}

\usepackage[T1]{fontenc}

\usepackage{pifont}

\DeclareRobustCommand{\VAN}[3]{#2}
\let\VANthebibliography\thebibliography
\def\thebibliography{\DeclareRobustCommand{\VAN}[3]{##3}\VANthebibliography}

\usepackage{graphicx}	% Including figure files
\usepackage{amsmath}	% Advanced maths commands

\title[X-ray weak AGN revealed by JWST and Chandra]{JWST meets Chandra: a large population of Compton thick, feedback-free, and intrinsically X-ray weak AGN, with a sprinkle of SNe}

\author[Maiolino et al.]{\parbox{\textwidth}{
Roberto Maiolino,$^{1,2,3}$\thanks{E-mail: rm665@cam.ac.uk}
Guido Risaliti,$^{4,5}$
Matilde Signorini,$^{5,6}$
Bartolomeo Trefoloni,$^{1,4,5}$
Ignas Juod\v{z}balis,$^{1,2}$
Jan Scholtz,$^{1,2}$
Hannah \"{U}bler,$^{1,2}$
Francesco D'Eugenio,$^{1,2}$
Stefano Carniani,$^{7}$
Andy Fabian,$^{8}$
Xihan Ji,$^{1,2}$
Giovanni Mazzolari,$^{9,10}$
Elena Bertola,$^{5}$
Marcella Brusa,$^{9,10}$
Andrew J. Bunker,$^{11}$
Stephane Charlot,$^{12}$
Andrea Comastri,$^{10}$
Giovanni Cresci,$^{5}$
Christa Noel DeCoursey,$^{13}$
Eiichi Egami,$^{13}$
Fabrizio Fiore,$^{14,15}$
Roberto Gilli,$^{10}$
Michele Perna,$^{16}$
Sandro Tacchella,$^{1,2}$
Giacomo Venturi$^{7}$
\\
}
\\
\parbox{\textwidth}{
$^{1}$Kavli Institute for Cosmology, University of Cambridge, Madingley Road, Cambridge, CB3 OHA, UK\\
$^{2}$Cavendish Laboratory - Astrophysics Group, University of Cambridge, 19 JJ Thomson Avenue, Cambridge, CB3 OHE, UK\\
$^{3}$Department of Physics and Astronomy, University College London, Gower Street, London WC1E 6BT, UK\\
$^{4}$Dipartimento di Fisica e Astronomia, Università di Firenze, via G. Sansone 1, 50019 Sesto Fiorentino, Firenze, Italy\\
$^{5}$INAF – Osservatorio Astrofisico di Arcetri, Largo Enrico Fermi 5, I-50125 Firenze, Italy\\
$^{6}$Dipartimento di Matematica e Fisica, Univeristà di Roma 3, Via della Vasca Navale, 84, 00146 Roma RM\\
$^{7}$Scuola Normale Superiore, Piazza dei Cavalieri 7, I-56126 Pisa, Italy\\
$^{8}$Institute of Astronomy, University of Cambridge, Madingley Road, Cambridge CB3 0HA, UK\\
$^{9}$Dipartimento di Fisica e Astronomia, Università di Bologna, Via Gobetti 93/2, I-40129 Bologna, Italy\\
$^{10}$INAF – Osservatorio di Astrofisica e Scienza dello Spazio di Bologna, Via Gobetti 93/3, I-40129 Bologna, Italy\\
$^{11}$Department of Physics, University of Oxford, Denys Wilkinson
Building, Keble Road, Oxford OX1 3RH, UK\\
$^{12}$Sorbonne Universit\'e, CNRS, UMR 7095, Institut d'Astrophysique de Paris, 98 bis bd Arago, 75014 Paris, France\\
$^{13}$Steward Observatory, University of Arizona, 933 N. Cherry Avenue, Tucson, AZ 85721, USA\\
$^{14}$INAF - Osservatorio Astronomico di Trieste, Via G. B. Tiepolo 11, I–34131 Trieste, Italy\\
$^{15}$IFPU - Institut for fundamental physics of the Universe, Via Beirut 2, 34014 Trieste, Italy\\
$^{16}$Centro de Astrobiolog\'ia (CAB), CSIC-INTA, Ctra. de Ajalvir km 4, Torrej\'on de Ardoz, E-28850, Madrid, Spain\\
}
}

\date{Accepted XXX. Received YYY; in original form ZZZ}

\pubyear{2015}

\begin{document}
\label{firstpage}
\pagerange{\pageref{firstpage}--\pageref{lastpage}}
\maketitle

\begin{abstract}
We investigate the X-ray properties of a  sample of 71 broad line and narrow line AGN at 2$<$z$<$11 discovered by JWST in the GOODS fields, which have the deepest Chandra observations ever obtained. Despite the widespread presence of AGN signatures in their rest-optical and -UV spectra, the vast majority of them is X-ray undetected. The stacked X-ray data of the non-detected sources also results in a non-detection. The upper limit on the X-ray emission for many of these AGN is one or even two orders of magnitude lower than expected from a standard AGN SED. X-ray absorption by clouds with large (Compton-thick) column density and low dust content, such as the Broad Line Region (BLR) clouds, can explain the X-ray weakness. In this scenario the BLR covering factor should be much larger than in low-z AGN or luminous quasars; this is supported by the larger equivalent width of the broad component of H$\alpha$ in JWST-selected AGN. We also find that the JWST-discovered AGN lack prominent, fast outflows, suggesting that, in JWST-selected AGN, dense gas lingers in the nuclear region, resulting in large covering factors. We also note that a large fraction of JWST-selected AGN matches the definition of NLSy1, typically accreting at high rates and characterized by a steep X-ray spectrum -- this can further contribute to their observed weakness at high-z. Finally, we discuss that the broad Balmer lines used to identify type 1 AGN cannot be ascribed to Very Massive Stars or Supernovae, although we show that some of the faintest broad lines could potentially be associated with superluminous SNe.
\end{abstract}


%%%%%%%%%%%%%%%%%%%%%%%%%%%%%%%%%%%%%%%%%%%%%%%%%%

%%%%%%%%%%%%%%%%% BODY OF PAPER %%%%%%%%%%%%%%%%%%

% {\bf \color{red} PLEASE DO NOT FORWARD OUTSIDE THE COLLABORATION, THANKS!}

\section{Introduction}

The launch of the James Webb Space Telescope (JWST) is drastically revisiting our understanding of black hole formation and accretion in the early Universe. Indeed, its unprecedented sensitivity in the infrared bands has allowed the discovery of several dozens Active Galactic Nuclei (AGN) at high redshift (z$\sim$3--11) with bolometric luminosities ($L_{AGN}\sim 10^{42}-10^{45}\rm~erg/s$)\footnote{Throughout the paper $L_{AGN}$ indicates the bolometric luminosity.} much lower than probed by previous surveys of high-z quasars.

Most of these AGN have been identified primarily through the detection of a broad component of the permitted emission lines (primarily H$\alpha$ and H$\beta$) without a corresponding broad component of forbidden lines (mostly [OIII]5007), indicating that it cannot be associated with an outflow and leaving the Broad Line Region around an accreting black hole (BH) as the most plausible explanation
\citep{Kocevski23,Ubler2023,Ubler24,Harikane_AGN,Maiolino23c,Maiolino24_GN-z11,Kokorev2023_AGN,Furtak2023_AGN,Greene2024,Juodzbalis2024,Kocevski2024}. 

Additionally, a growing number of narrow-line, type 2 AGN has also been identified by using various diagnostics \citep{Scholtz23,Chisholm2024,Perna2023,Goulding2023_AGN,Bogdan23} and new diagnostic diagrams are being developed to identify type 2 AGN \citep[e.g.][]{Mazzolari24}.

This new discovery space has revealed important and interesting features of the population of early super-massive BHs. They tend to be overmassive relative to the host galaxy stellar mass, when compared with the local relation \citep{Ubler2023,Harikane_AGN,Maiolino23c,Maiolino24_GN-z11,Bogdan23,Goulding2023_AGN,Juodzbalis2024,Parlanti2024ALESS}, indicating that BH formation and growth outpaces star formation in the host galaxy. Some studies have suggested that the observed overmassive nature of these black holes is due to a very large scatter of the $M_{BH}-M_{star}$ relation combined with selection effects \citep{Li_Silvermann_2024}.  However, high-z BHs are much closer to the local BH-$\sigma$ relation \citep{Maiolino23c,Juodzbalis2024} suggesting that the latter relation is more fundamental and universal, and also that selection effects play a secondary role. The BHs discovered by JWST have been found to span a broad range of accretion rates, from super-Eddington to orders of magnitude below the Eddington limit. Generally, the results obtained by JWST have been interpreted as requiring either massive seeds (e.g. Direct Collapse Black Holes) and/or black holes experiencing bursts of super-Eddington accretion \citep{Schneider2023,Trinca2023,Jeon2024,Bennett2024,Zhang_Trinity_23_I,DiMatteo2023,Scoggins2023,Lupi2023,Rantala2024,Tremmel2023,Natarajan2024,Pacucci2023,Volonteri2023}

The fraction and density of AGN found by JWST depends on the selection criteria and on the depth of the observations. 
\cite{Greene2024} selected AGN through their red colours; out of a population of red objects accounting for about 1 percent of the galaxy population, they find that most of them are spectroscopically confirmed as AGN.
\cite{Matthee2023} selected AGN from NIRCam grism slitless spectra based on the detection of broad H$\alpha$; they derive that, about 1\% of galaxies in the UV luminosity range $-18<M_{UV}<-20$ host a type 1
AGN with bolometric luminosity higher than $10^{44.5}~erg/s$ at z$\sim$4--5. \cite{Harikane_AGN} selected type 1 AGN at lower bolometric luminosities by using NIRSpec spectroscopic data finding that the fraction of type 1 AGN at z$\sim$4-7 ranges from a few percent to $\sim$10-15\%. \cite{Maiolino23c} used deep spectroscopic data from the JWST Advanced Deep Extragalactic Survey \citep[JADES][]{Eisenstein23} survey to identify AGN at z$>$4, and found that, in the UV luminosity range $-18<M_{UV}<-20$, about 10\% of galaxies host a type 1 AGN with $L_{AGN}>10^{44}~erg/s$. \cite{Scholtz23} used a smaller sample of the JADES spectra to identify also type 2 AGN, and found that about 20\% (in the UV luminosity range $-18<M_{UV}<-20$) host either a type 2 or a type 1 AGN at z$\sim$5, down to bolometric luminosities of $\sim 10^{42}~erg/s$.

These space densities are much higher than the extrapolation of the luminosity function of luminous quasars \citep{Niida2020}, indicating that JWST is tracing a different population, which is not surprising. However, even more intriguingly, the population of AGN found by JWST is much larger, by about an order of magnitude, than X-ray selected AGN at similar redshifts \citep{Giallongo19}. Yet, X-ray selected AGN resolved about 90\% of the X-ray background (XRB) in the 0.5--2~keV and 2--10~keV bands \citep{Moretti2003}. Therefore, if the AGN discovered by JWST share the same SED as AGN identified locally and at lower redshift, then there is not much scope for an additional large population of AGN within the constraints given by the XRB. This issue was indeed highlighted by \cite{Padmanabhan2023}, who pointed out that the population of AGN discovered by JWST would overproduce the XRB by up to an order of magnitude. However, they assumed a SED inferred from local/low-z AGN and QSOs \citep{Shen2020}. Therefore, the main issue is whether this new population of AGN discovered by JWST shares the same observed SED as local/low-z AGN and, in particular, the same observed X-ray to optical/UV luminosity ratio. Some studies have already noticed that the JWST-identified AGN at high redshift are undetected in the X-rays \citep[e.g.][]{Ubler2023,Matthee2023}.
X-ray weakness at high-z was found also in some AGN not selected with JWST, such as the red quasars at z$\sim$2--3 found by \cite{Ma2024RedQSOs} and the red quasar at at z=7.2 discovered by \cite{Fujimoto2022}; however, JWST is finding a larger AGN population with these properties than previous studies. For instance,
more recently, \cite{Yue2024}  and \cite{Ananna2024} stacked X-ray data of the so-called Little Red Dots {\citep{Greene2024,Kocevski2024}}, many of which are found to host an AGN, and obtained non-detections or marginal detections. However, the Little Red Dots are only a small fraction ($\sim$10\%--30\%) of the AGN population found by JWST \citep{Hainline2024_LRD}. A more systemic characterization of the X-ray properties of the bulk of the AGN population discovered by JWST at high-z, has yet to be obtained.

Here we explore the X-ray properties of the AGN discovered by JWST in the deepest Chandra fields, GOODS-N (2 Msec) and GOODS-S (7 Msec) \citep{CDFN-cat,CDFS-cat} and demonstrate that they are much fainter in the hard X-rays than expected from the standard SED of AGN, even accounting for its luminosity dependence. We discuss possible scenarios to explain such differences, ranging from heavy obscuration (Compton thick), intrinsic X-ray weakness (associated with high accretion rate and/or lack of a hot corona), and, in a small number of cases, the possible misclassification. We show that there is observational evidence in support of some of these scenarios.

Throughout this work, we use the AB magnitude system and assume a flat $\Lambda$CDM cosmology with $\Omega_m=0.315$ and $H_0=67.4$ km/s/Mpc \citep{Planck20}.  With this cosmology, $1''$ corresponds to a transverse distance of 5.84 proper kpc at $z=6$.

\section{JWST sample of AGN and bolometric luminosities}

We focus on AGN discovered (or further characterised) by JWST in the GOODS-N and GOODS-S fields. These are the deepest Chandra fields and therefore, together with the deep JWST spectroscopic data, provide some of the most stringent constraints on the X-ray emission.
The primary focus is at z$>$4, the new AGN frontier explored by JWST at intermediate/low luminosities. However, we consider also some samples at 2$<$z$<$4 to explore whether some of the observed properties are also seen at intermediate redshifts.
More specifically, we consider the following samples:

\begin{itemize}

    \item The type 1 AGN at 4$<$z$<$5.6 found by \cite{Matthee2023} in GOODS-S and GOODS-N through the detection of a broad component of H$\alpha$ in the NIRCam grism slitless spectra from the FRESCO survey \citep{Oesch2023} (eight objects in total). In most of these cases, there are no [OIII] observations to compare the profile and verify the lack of a broad component. However, the width and strength of the emission lines, together with point-like morphologies, leave little doubt that these are type 1 AGN. Additionally, most of these sources have very red colors in the reddest NIRCam bands, suggesting that they are absorbed with $A_V\sim 1-4$~mag.
    
    \item The type 1 AGN discovered with NIRSpec-MSA \citep{Jakobsen22,Ferruit22,Boker23} observations in GOODS-N and GOODS-S at 4$<$z$<$11 as part of the JADES survey \citep{Eisenstein_Jades} from \cite{Maiolino23c} (13 objects in total). These are selected through the detection of a broad component of H$\alpha$ (and in one case H$\beta$), without a corresponding broad component of [OIII]5007. The only exception is GN-z11, for which the detection of a type 1 AGN is identified through densities typical of the BLR, via the UV semi-forbidden lines of NIII] and NIV], and for which the presence of the AGN is also confirmed by other AGN-like transitions (CII$^*$ and [NeIV]), fast outflows revealed in CIV \citep{Maiolino24_GN-z11}, ionization cones \citep{Maiolino23b} and a large Ly$\alpha$ halo typical of high-z quasars \citep{Scholtz2023_GN-z11}. The X-ray non-detection of GN-z11 was already discussed in \cite{Maiolino24_GN-z11}, and the upper limits were pointed out to be consistent with the Narrow Line Sy1 nature of GN-z11.
    
    \item The type 1 AGN GS\_3073 at z$=$5.5 in GOODS-S. This was originally considered to be an AGN by \cite{vanzella+2010} and \cite{grazian+2020} based on the detection of NV and NIV] in the UV rest-frame, with ground-based spectroscopy. It was then found to be a type 1.8 AGN by \cite{Ubler2023} based on the detection of broad components of multiple permitted lines (H, HeI, and HeII) in NIRSpec-IFS \citep{Boker2022} data. \cite{Ubler2023} and \cite{Ji2024} detected also several additional high ionization transitions, including coronal lines. The lack of X-ray detection of this AGN was already mentioned in the papers above, but will be quantified more in detail in this paper.
    
    \item The dormant black hole at z=6.7 found by \cite{Juodzbalis2024} in JADES, through the detection of a broad component of H$\alpha$.
    
    \item We also include XID403, an X-ray selected, Compton-thick AGN at z$=$4.76 in GOODS-S initially discovered by \cite{Gilli2011}. This galaxy has been recently observed with NIRSpec-IFS revealing a clear broad component of the H$\alpha$ line, indicating that this is actually a reddened type 1 AGN, whose X-rays are heavily absorbed by a Compton thick medium
    \citep{Parlanti2024ALESS}.

    \item We include six additional type 1 AGN at 2$<$z$<$4 found in the public DR3 of JADES \citep{DEugenio2024}. One of these is characterized by deep, slightly blueshifted absorption of H$\alpha$, H$\beta$, and HeI (Juod\v{z}balis et al. in prep.). 
    
    \item We also consider the sample of 41 type 2, narrow line AGN found by \cite{Scholtz23} at 2$<$z$<$9 based on NIRSpec-MSA spectra from JADES. As the classical type 2 classification based on BPT diagrams breaks down at high redshift for intermediate luminosity and low metallicity AGN \citep{Kocevski_AGN,Ubler2023,
    Maiolino23c},  \cite{Scholtz23} identified type 2 AGN also based on the detection of high ionization transitions and diagnostic diagrams based on UV transitions. The sample of \cite{Scholtz23} also includes GS-21150, which was allocated a slit in the JADES MSA Deep survey, because X-ray detected; this galaxy was identified spectroscopically to have a redshift of 3.08, however none of the optical or UV diagnostics indicate that it hosts an AGN.
    
    \item Finally, we include the type 2 AGN at z=5.6 found by \cite{Chisholm2024} via the detection of [NeV] with NIRSpec-MSA.
\end{itemize}

\begin{figure}
	\centering
	\includegraphics[width=0.5\linewidth]{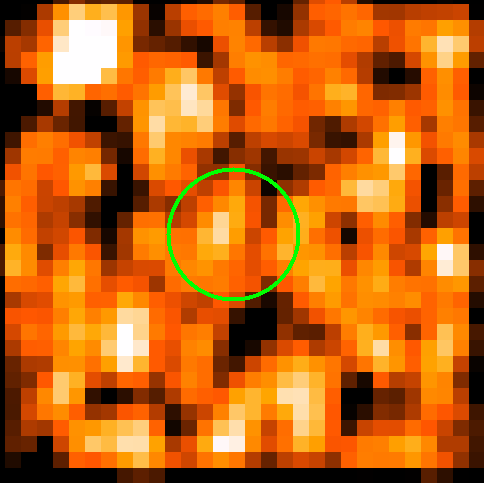}
	 \caption{{\em Chandra} X-ray image of the type 1.8 AGN GS\_3073,
 which is one the most luminous z$>$4 targets in our sample, thoroughly studied \citep{Ubler2023,Ji2024}, and whose AGN nature is unambiguously
   confirmed by numerous rest-optical and rest-UV diagnostics. The image does not reveal a detection, while a $\sim$20$\sigma$ detection would have been expected based on a standard AGN SED. The radius of the circular region is 3 arcsec.}
	\label{fig:GS3073}
\end{figure}

Overall, type 1 AGN are primarily identified through the identification of a broad component of H$\alpha$ (and in some cases of H$\beta$ and other permitted lines). However, the sensitivity in detecting broad lines is different in the various surveys, with the slitless grism surveys being the least sensitive, while the JADES slit survey being the deepest. Additionally, the parent samples are heterogeneous: the slitless surveys provide an unbiased sample of broad line AGN down to their sensitivity limit; multi-slit (MSA) spectroscopy has, for different surveys, a complex selection function, primarily based on multi-band photometric redshifts and a limiting magnitude to ensure continuum or (narrow) lines detection, but generally without a-priori knowledge for the presence of an AGN. 

The type 2 AGN sample is selected with a variety of narrow line diagnostics, primarily consisting of the detection of high ionization lines and UV/optical diagnostics diagrams calibrated with models or empirically. The parent sample of the large type 2 sample from \cite{Scholtz23} has the same heterogeneity as discussed above.

Despite the heterogeneous diagnostics of the parent samples, {\it all these AGN have in common the fact that they probe a low/intermediate luminosity regime at high redshift, not probed by previous samples, and that were not pre-selected because known to have an AGN from previous observations} (with the exception of the two X-ray selected objects discussed above). Additionally, these JWST-idenfied AGN are typically hosted in low mass ($M_*\sim 10^8-10^{10}~M_\odot$) and metal poor ($Z\sim 0.1~Z_\odot$) host galaxies, in contrast with more luminous quasars at similar redshifts.

The bolometric luminosities of type 1 AGN were not inferred from the continuum \citep[with the exception of GN-z11, see ][]{Maiolino24_GN-z11} as it may be dominated by the host galaxy or by reflected light. They are instead homogeneously estimated from the extinction-corrected broad component of H$\alpha$ and adopting the scaling relation provided by \cite{SternLaor2012}. For type 2 AGN the bolometric luminosity is inferred from the extinction-corrected [OIII]5007 line and adopting the scaling relation given in \cite{Scholtz23}, based on the photoionization models obtained by Hirschmann et al. (in prep.). We will however also illustrate the effect of using the scaling relations provided by \cite{Netzer19}, which typically give luminosities that are 0.5 dex higher.

The full sample of AGN, with their primary properties, is given in Tables \ref{tab:type1s} and \ref{tab:type2s}, for type 1 and type 2 respectively.

\begin{figure}
\includegraphics[width=1.0\linewidth]{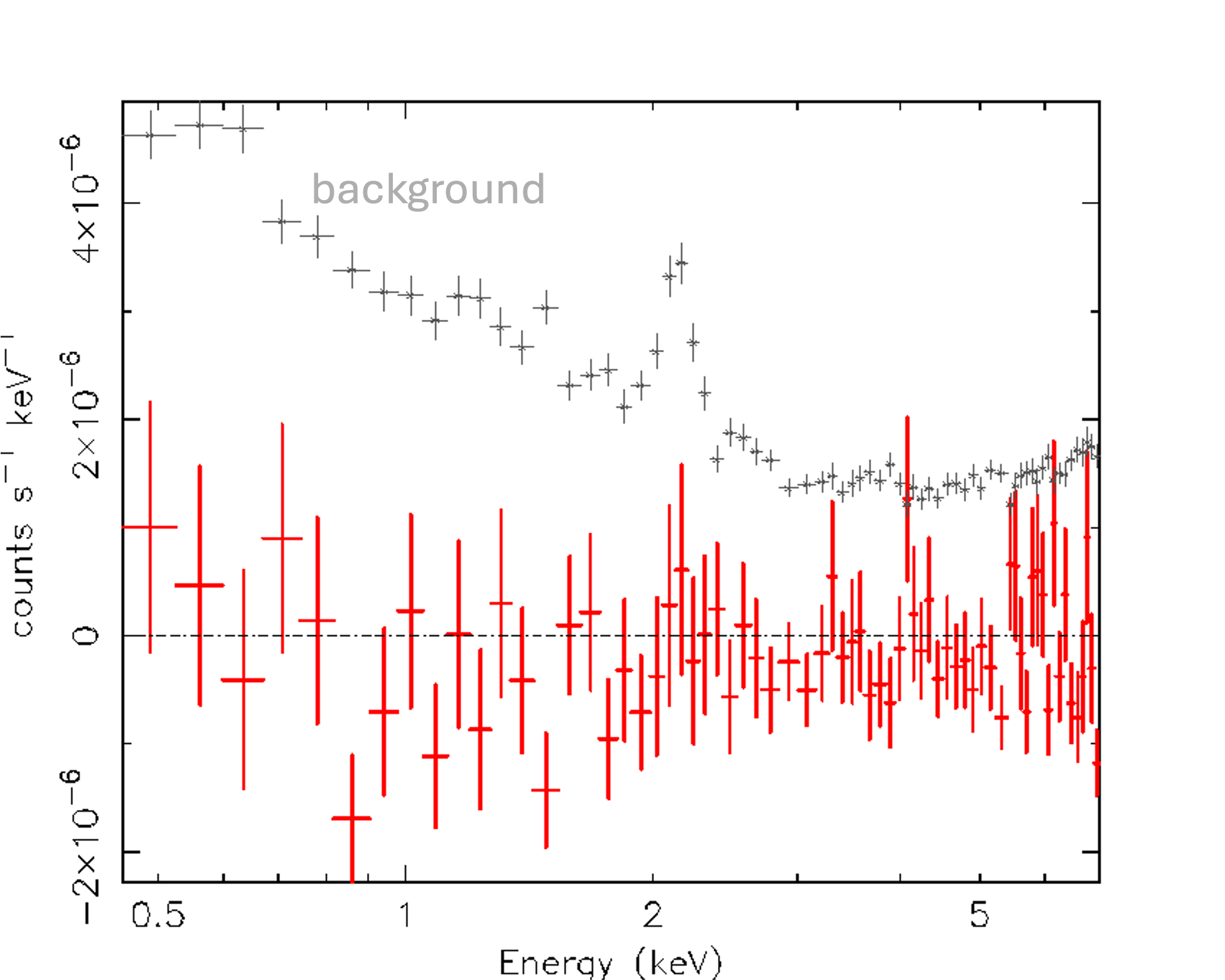}
	\caption{Stacked X-ray spectrum for the whole sample of JWST-identified type 1 AGN. The upper grey spectrum shows the background level.} 
	\label{fig: Xspectra}
\end{figure}

\begin{figure*}
	\centering
\includegraphics[width=0.7\linewidth]{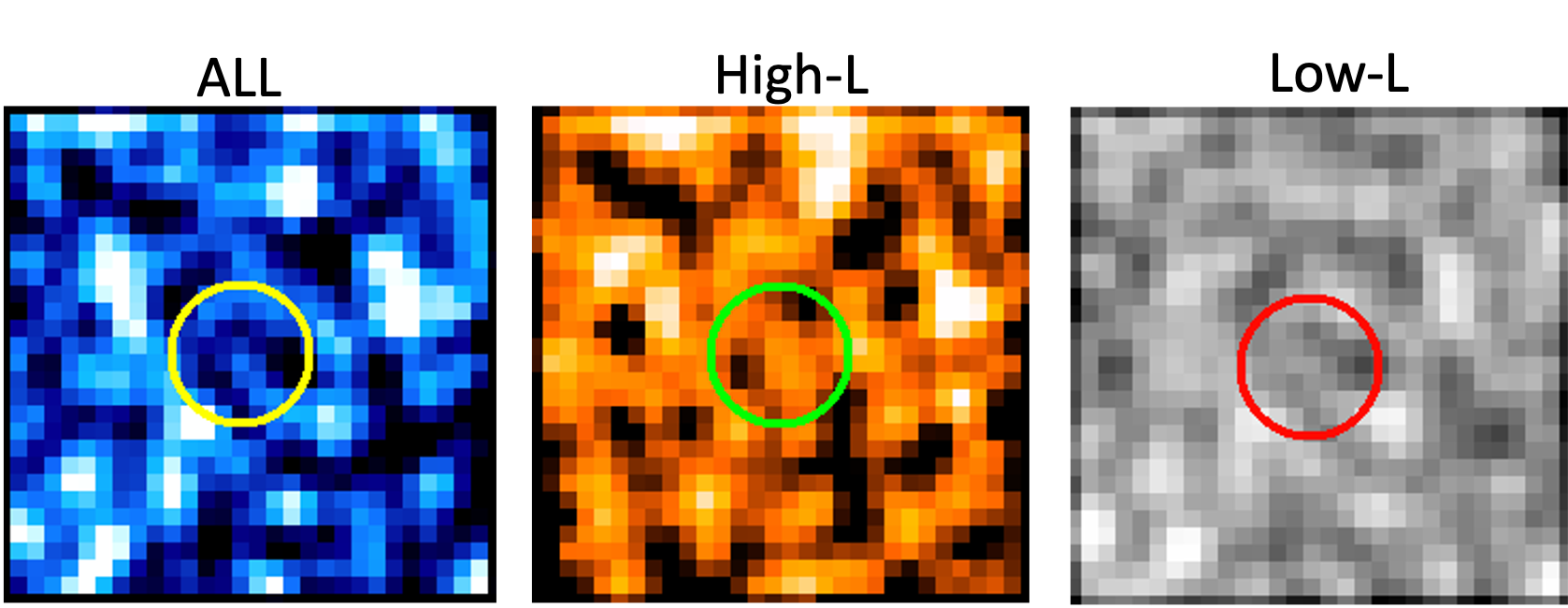}
	 \caption{Stacked {\em Chandra} X-ray images of the whole type 1 AGN sample (left 
   panel) and for the high-luminosity and low-luminosity subsamples (middle and right panels, respectively). The radius of the circular region is 3 arcsec. The individual images are from different positions in the CDF-N and CDF-S fields, so they are shown for illustrative purposes only. All the fluxes have been derived from a spectral analysis of each individual
  source, as described in the text.}
	\label{fig: thumbnails1}
\end{figure*}

\section{X-ray data}

All the sources discussed in this paper are in either the Chandra Deep Field North (hereafter CDF-N) or the Chandra Deep Field South (hereafter CDF-S).
CDF-N and CDF-S are the two deepest X-ray surveys to date, with a total exposure of 2 Msec and 7 Msec, respectively, and an on-axis flux limit in the 0.5-2.0~keV band of 2.5$\times10^{-17}$~erg cm$^{-2}$ s$^{-1}$ (CDF-N, \citealt{CDFN-cat}) and 6.4$\times10^{-18}$~erg cm$^{-2}$ s$^{-1}$ (CDF-S, \citealt{CDFS-cat}). This on-axis sensitivity significantly decreases for sources in the outer regions of the fields (see \citealt{CDFN-cat} and \citealt{CDFS-cat} for details).

The total exposure in both fields has been achieved through many individual observations (102 for the CDF-S and 20 for the CDF-N). Since each source is located in different positions during each individual observations, calibrations and PSF change for each observation. Therefore, in order to obtain accurate spectral information (in particular, to correctly estimate flux upper limits) we extracted and calibrated the X-ray spectral counts (source + background) from each individual Chandra exposure. This was done by using the CIAO 4.16 package \citep{CIAO}, and regardless of the presence (or absence) of a detected source. Since in almost all cases the source is not detected, this amounts to extracting a background spectrum. The optimal radius of the circular extraction region depends on the source position (due to the highly variable PSF across the field of view of Chandra), and has been chosen in order to include 90\% of the effective area at that position. The "true" background has been extracted from an annular region with the same center, paying attention to remove any serendipitous source present in the chosen annulus. Calibrations were computed for individual exposures too. The individual spectral data and calibrations have then been stacked to produce a total source and background spectrum. 

The evaluation of the flux upper limits has been performed through a spectral analysis\footnote{With "spectral analysis" here we mean a procedure that follows for our non-detections the same steps as a standard spectral analysis, formally treating the counts (source+background) at the source positions as spectra. The upper limits are obtained from the error estimate of the flux for this (background-subtracted) component.} with the XSPEC 12.6 package \citep{XSPEC}. 
We assumed a power law model with a fixed photon index $\Gamma$=1.7 and a Galactic absorption. The photon index $\Gamma$=1.7 has been chosen in order to test the scenario of ``standard'' AGN, which are typically characterized by this X-ray spectral slope.
Having fixed the photon index and Galactic absorption, the only free parameter was the source flux. In order to derive such flux, we formally fitted the data in the 0.5-5~keV (observed frame) interval, corresponding to a rest-frame energy range of 2.5-25~keV at z=4.  We considered the source as detected if the flux best fit value is higher than zero with a statistical significance of at least 3~$\sigma$. In case of non detections, we computed the 90\% upper limit with respect to the best fit value. The whole procedure has been performed using the C-statistics, which correctly accounts for Poissonian counts distributions. The spectral data has not been grouped. 

Fluxes and upper limits have been estimated in the 2-10~keV energy range both in the observed frame and in the rest frame (in this case we used an extension of the instrumental response to energies slightly lower than 0.5~keV when needed).  Luminosities have been estimated assuming a flat $\Lambda$CDM cosmology with $\Omega_M=0.7, \Omega_\Lambda$=0.3. 

%In the case of GN-z11, which is clearly a Narrow Line Seyfert 1, we will also provide the upper limit using the steeper slope typical of NLSy1, for consistency with \cite{Maiolino24_GN-z11}. 
These values are reported in Tables~\ref{tab:type1s} and \ref{tab:type2s} for type~1 and type~2 objects, respectively.  We have checked that overall our estimates are consistent with the {\em Chandra} upper-limit maps of \citet{Xue16} and \citealt{CDFS-cat}.

We note that while for most of the sources the net counts after background subtraction are consistent with zero within 1~$\sigma$, in a few cases they are above zero with a statistical significance between 1 and 2~$\sigma$. We treat these cases as non-detections, considering that our upper limits estimates automatically take into account the level of these possible marginal detections. Indeed, the two sources with a signal-to-noise ratio above zero at the $\sim$1.5-2~$\sigma$ level, GN 3608 and GN 73488, have the two highest flux upper limits in our sample (Table~\ref{tab:type1s}). 

The images shown throughout the paper have been extracted from the merged images of both fields and are intended for visual inspection only.

As is clear from Tables~\ref{tab:type1s} and \ref{tab:type2s}, the vast majority of the JWST-identified AGN are undetected with Chandra. Out of the five detected sources, two had been originally selected in the X-rays, and will be discussed in the next Sections. The remaining three objects, GN~721, GS~49729 and GS~209777, are type 1 AGN at redshift z$\sim$3 (Table~\ref{tab:type1s}). GN~721 is detected with a statistical significance of $\sim5~\sigma$, so only a basic spectral analysis is possible.  We repeated the spectral analysis for these objects allowing for a free photon index, finding a low value ($\Gamma=0.1\pm0.6$) suggesting a Compton-reflection spectrum. The bolometric to X-ray ratio for this source is $\sim$10 times higher than in typical optically selected AGN (see discussion in the next section). This is of the same order as the ratio between the flux of the primary X-ray power law and the Compton-reflection component in local AGN in the same rest-frame interval (1.5-15 keV). Therefore, the X-ray weakness of GN~721 could be interpreted as due to absorption and reflection by a "standard" Compton-thick circumnuclear torus of an otherwise typical intrinsic X-ray emission. The remaining two sources, GS~49729 and GS~209777, have high signal-to-noise detections in the CDF-S field, and show X-ray properties typical of a "normal" AGN: their spectra are well reproduced by an unobsorbed power law with a photon index $\Gamma=1.65\pm0.05$ and $\Gamma=1.7\pm0.1$, respectively, and their bolometric to X-ray flux ratios are typical of optically selected AGN (see next section).    

 Considering that the low photon index found in GN~721 may be common among our sources, we repeated the upper limits estimates assuming a value $\Gamma=0$ for the whole sample. While the individual flux upper limits may change within a factor of two, the overall distribution does not significantly change. Therefore, in the remainder of the paper we will always refer to the upper limits obtained assuming $\Gamma=1.7$.

In order to increase the sensitivity of our analysis, we also produced stacks in luminosity bins, which also give non-detections, and yield upper limits down to a few times $10^{-18}\rm~erg/s/cm^2$ and a few times $10^{-19}\rm~erg/s/cm^2$, for the Type 1 and Type 2 samples, respectively. Note that the upper limit on the type 2 sample is more stringent both because there are more type 2 AGN and because the analysis performed by \cite{Scholtz23}, out of which our type 2 sample is extracted, is done only in GOODS-S, where the Chandra observation is deeper.

We do not show the images and spectra for all non detections. We only show, as an example, the image for the remarkable case of the type 1.8 AGN GS\_3073. As we will discuss further later on, this is a fairly luminous AGN with multiple unambiguous UV and optical AGN signatures. As illustrated in Fig.~\ref{fig:GS3073}, despite being in the deepest region of the CDFS 7Ms area, it is not detected.

\begin{figure}
        \centerline{
	\includegraphics[width=0.5\linewidth]{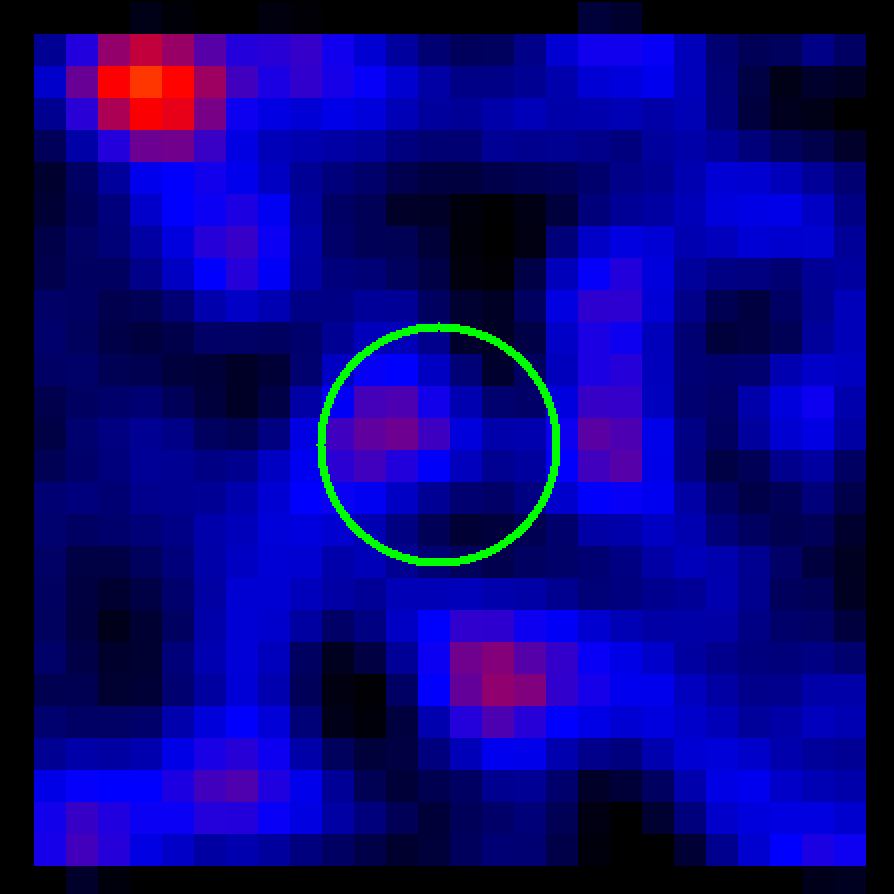}}
	\caption{Stacked {\em Chandra} X-ray images of the whole type 2 AGN sample. The radius of the circular region is 3 arcsec.} 
	\label{fig: thumbnail2}
\end{figure}

Fig.~\ref{fig: Xspectra} shows the X-ray spectrum obtained for the stack of all type 1 AGN found by JWST in the GOODS fields, which shows no detection. We have also stacked this sample after splitting it in two bins of high and low luminosities, with a dividing value of $\log(L_{BOL})=44.8$, consisting of 11 and 12 sources, respectively. These stacks are shown in the Appendix, and do not show any detection either. The thumbnails associated with these stacks are shown in Fig.~\ref{fig: thumbnails1}. The stacking of type 2 AGN is shown in Fig.~\ref{fig: thumbnail2}, also in this case not showing any detection.

\begin{figure*}
	\includegraphics[width=0.8\linewidth]{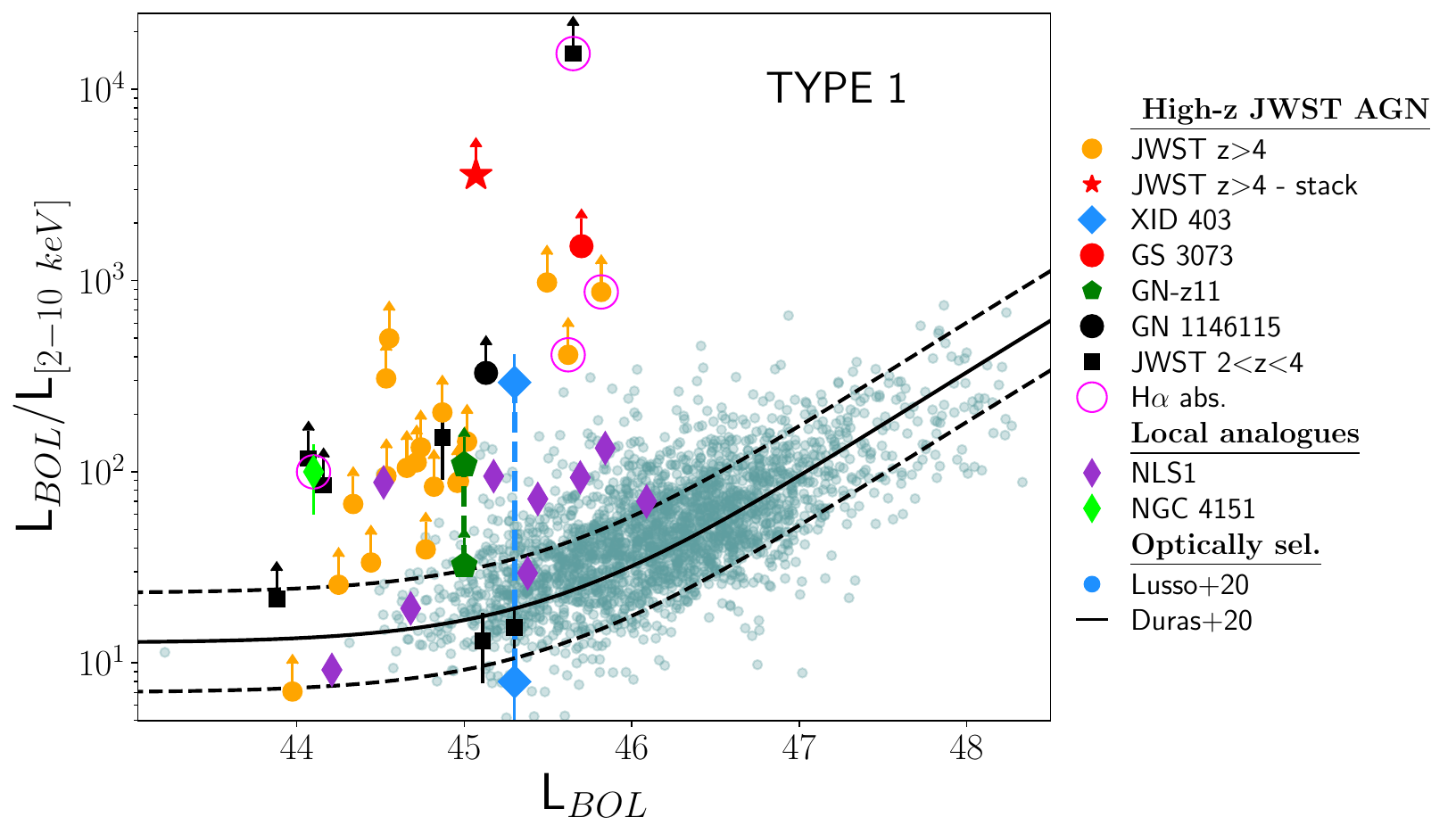}
	\caption{
 Ratio $k_{\rm bol,X}$ between the AGN bolometric luminosity and the X-ray (2-10~keV) luminosity as a function of the bolometric luminosity L$_{BOL}$ for our sample of type 1 AGN. Light blue small points are from the low-redshift sample of 
 \citet{Lusso2020}, 
 and are representative of ``normal'', optically/UV selected blue quasars. 
 The light black lines show the best-fit (continuous line) and dispersion (dashed lines) of the $k_{\rm bol,X}$-L$_{BOL}$ relation for quasars from \citet{Duras2020}. Purple diamonds show local NLSy1  while the light green diamond shows the prototypical broad line Seyfert 1 NGC~4151.
    Orange solid circles show all lower limits in $k_{\rm bol,X}$ for our sample of type 1 AGN at z$>$4.  The star shows the lower limit for the stacked spectra from the whole sample at z$>$4. We also identify a few relevant sources, as shown in the figure legend (see text). The black squares show a group of lower redshift (2$<$z$<$4) type 1 AGN in JADES, two of which are detected.
    Sources with indication of absorption features in their broad H$\alpha$ profile are highlighted with blue empty circles.
    The bottom and top points relative to the source XID 403 refer to the observed and intrinsic X-ray luminosity, respectively. 
    The bottom and top points relative to GN-z11 indicate two different ways of estimating its X-ray upper limit (see text).
    }
	\label{fig:Rbol_type1}
\end{figure*}

\begin{figure}
	\includegraphics[width=1.0\linewidth]{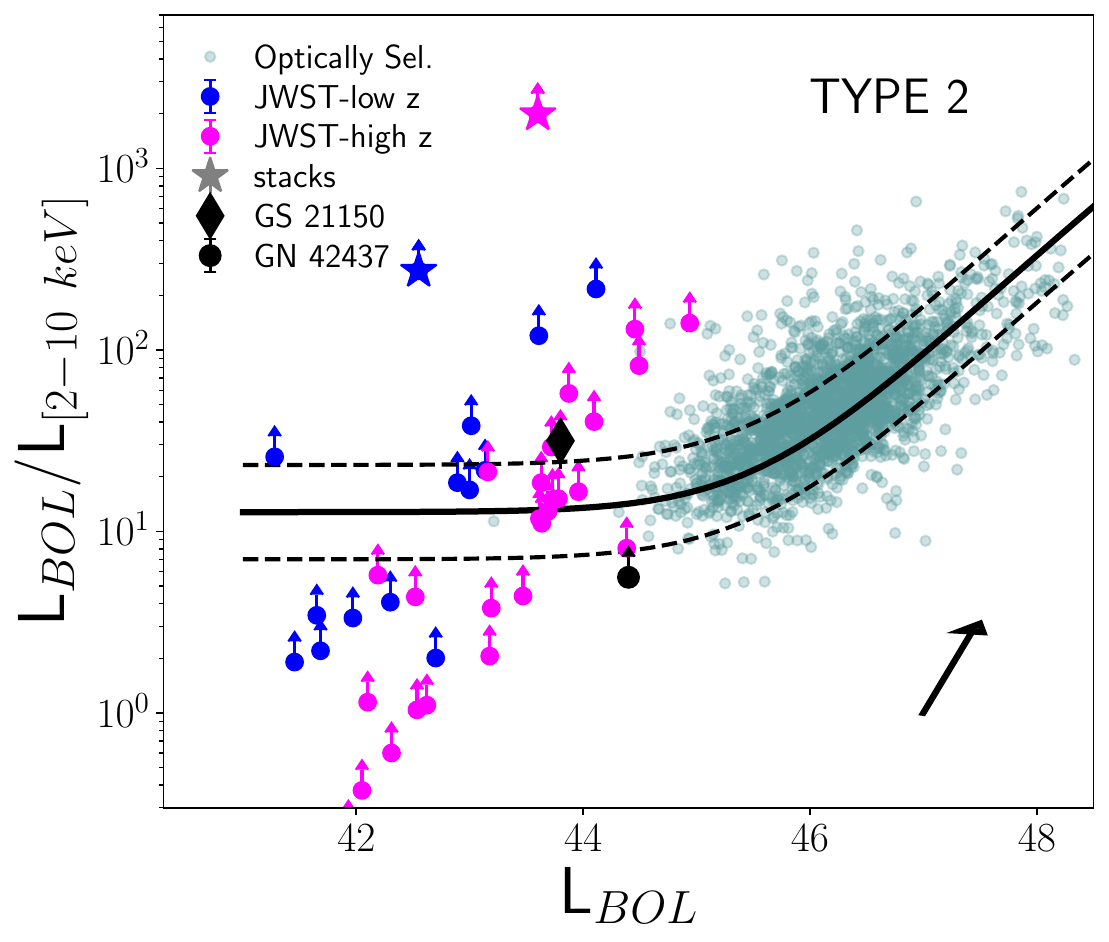}
	\caption{Same as Fig.\ref{fig:Rbol_type1} but for the type 2 sample. Blue and magenta points represent sources at redshift $z<3$ and $z>3$, respectively. Stars show the values for the stacked spectra obtained from the two subsamples. The two sources GS 21150 (the only X-ray detected object in our sample) and GN 42437 (the only type 2 not in the JADES sample) and are shown with different symbols. The arrow on the bottom-right corner of the plot shows the shift of the positions of the points if the bolometric luminosities are estimated as described in \citet{Netzer19}  (using their relation between narrow-H$\beta$ and bolometric luminosity) instead of as in \citet{Scholtz23}.}
    \label{fig:Rbol_type2}
\end{figure}

\section{Extremely high \texorpdfstring{$L_{Bol}/L_X$}{Lbol/LX}}
\label{sec:kbol}

The previous section has shown that the vast majority of AGN identified by JWST, both type 1 and type 2, are not detected even with the deepest Chandra observations, not even in stacking. A similar result was obtained, at lower redshifts, by \cite{Lyu2023} when exploring the X-ray counterparts of AGN photometrically identified in the mid-IR.

The only X-ray detected AGN at z$>$4 in our sample is XID403. This is primarily because it was initially selected based on its X-ray emission and subsequently observed with  JWST.
In the type 1 AGN sample at z$<$4 there are two X-ray detections out of six objects. 
There is only one type 2 AGN in the sample by \cite{Scholtz23} that is X-ray detected, at z=3.08; however, we note that this source was allocated a slit in the JADES survey because previously known to be X-ray detected, but its JWST spectrum does not reveal any AGN signature.

As far as we know, there is currently only one other
high-z AGN spectroscopically identified by JWST that is detected in the X-rays, UHZ-1 at z=10.1, identified by \cite{Bogdan23} and \cite{Goulding2023_AGN}. As XID403, UHZ-1 is a candidate Compton thick source, based on its X-ray spectrum. Its JWST spectrum does not reveal any AGN signature, further revealing the frequent disconnect between X-ray and optical/UV AGN diagnostics.

In this section, we investigate whether these non-detections, even in stacking, are inconsistent with standard AGN SED or if they are merely a result of the sensitivity limits of our observations, instead.

The key aspect to assess is whether the X-ray emission expected from the AGN bolometric luminosity is consistent with the X-ray upper limits or not. It is well known that AGN become X-ray weaker at high (quasar-like) bolometric luminosities \citep[e.g.][]{Duras2020,Shen2020}. Hence, it is important to compare the JWST-detected AGN with the SED inferred for low-z/local AGN with the same bolometric luminosity. Fig.~\ref{fig:Rbol_type1} shows the distribution of optically/UV selected type 1 AGN at z$<$0.6 from \citet{Lusso2020} (light blue small symbols) on a diagram where bolometric to hard X-ray luminosity ratio
($k_{bol,X} = L_{bol}/L_{[2-10~keV]}$) is plotted as a function of bolometric luminosity. The black line shows the best fit relation for local/low-z quasars and AGN identified by \citet{Duras2020}. 
We note that luminous (optically/UV selected) quasars at z$>$6 follow the same relation as the local/low-z AGN \citep{Zappacosta2023}, or possibly slightly above the relation in terms of $k_{bol,X}$ \citep{Vito2019}.

The large solid symbols indicate the values (mostly lower limits) inferred for the type 1 AGN identified by JWST in our GOODS-N/S sample. 

The only detection in the z$>$4 sample is the Compton thick AGN XID403, indicated by the upper blue diamond, while the lower diamond (connected with a dashed segment) indicates the X-ray absorption corrected value. It illustrates that the observed $L_{\rm Bol}/L_{\rm X}$ is well above the local, standard relation, while the corrected value is perfectly consistent with the relation. All other z$>$4 objects are lower limits, most of which are highly significant, i.e. well above the relation and its dispersion. Particularly luminous AGN which lacked X-ray detection, and which were individually discussed in the previous section, are marked with specific symbols (see legend) and are also well above the local relation. Remarkable is GS\_3073, which, as we mentioned, is an extremely well studied AGN at z=5.55 with multiple unambiguous AGN signatures, located in the deepest region of the Chandra 7 Ms pointing and, despite this, undetected and more than one order of magnitude above the standard relation.

The stacked X-ray limit for the z$>$4 type 1 sample is shown with a starred red symbol,  for the total sample (we do not plot the two stacks in two luminosity for the sake of clarity of an already overcrowded plot, but they convey the same information as the total stack, i.e. very high $L_{bol}/L_X$). These stacks provide some of the tightest constraints on the bolometric to X-ray ratio, deviating by more than two orders of magnitude relative to the local relation.

The black squares show the measurements for the (smaller) type 1 sample at 2$<$z$<$4. Most of them are upper limits, and most of these well above the local relation. In this case, as discussed in the previous section, there are three detections: GN 721 is well above the local relation, and is characterised by an X-ray spectrum suggestive of Compton thickness; GS 49729 and GS~209777 are consistent with the local relation, and their X-ray spectrum is consistent with a typical type 1 AGN without absorption.

Fig.~\ref{fig:Rbol_type2} shows the same diagram for the type 2 AGN in \cite{Scholtz23}. The black diamond shows the only X-ray detected galaxy (GS 21150, which was actually included in the sample because X-ray selected, but without AGN spectral signatures). All other type 2 AGN have lower limits on $L_{bol}/L_X$. Most of these individual lower limits are less constraining relative to the type 1 AGN. However, the stacked values (starred symbols) are above the local relation by one or two orders of magnitude, for the low-z (blue) and high-z  (magenta) bins, respectively. We also indicate with an arrow the effect of using the narrow H$\beta$ to bolometric correction provided by \cite{Netzer19}, which would make the deviations even stronger.

Summarizing, the X-ray analysis of the new population of high-z AGN identified by JWST, which have low/intermediate luminosities, reveals that they are significantly underluminous in X-rays compared to the lower redshift AGN population, or to the high-z population of luminous quasars. Specifically, for most of them, the X-ray emission is weaker by one or two orders of magnitude than expected based on their bolometric luminosity and assuming a local/standard AGN SED.

\section{Scenarios}

We now discuss the possible scenarios that could explain the X-ray weakness of the JWST-discovered AGN. There are three main scenarios: heavy absorption in the X-rays; intrinsic X-ray weakness; and misidentification of AGN. We anticipate that the latter case is very unlikely to play a significant role, as the same X-ray weakness is found independently of the AGN selection and identification method (broad lines, narrow line diagnostics, mid-IR excess). Additionally, there are several cases, such as GS\_3073, which are AGN beyond any doubt (strong broad permitted H and He lines, high ionization lines with high EW, and even coronal lines), which are extremely X-ray weak. However, we will discuss that some contamination for a minority of cases is potentially possible.

Most probably, a combination of all the effects discussed above is responsible for the X-ray weakness. 
However, we will focus primarily on the heavy obscuration scenario, as there is observational evidence that this is likely the case for a significant fraction of the sources.

%Clearly, 
Independently of the origin of the X-ray weakness, our finding that the bulk of the JWST-discovered AGN are extremely X-ray weak, i.e. they do not share the same SED as classical AGN templates, explains why their high abundance does not violate the X-ray background.

\begin{figure*}
    \includegraphics[width=\linewidth]{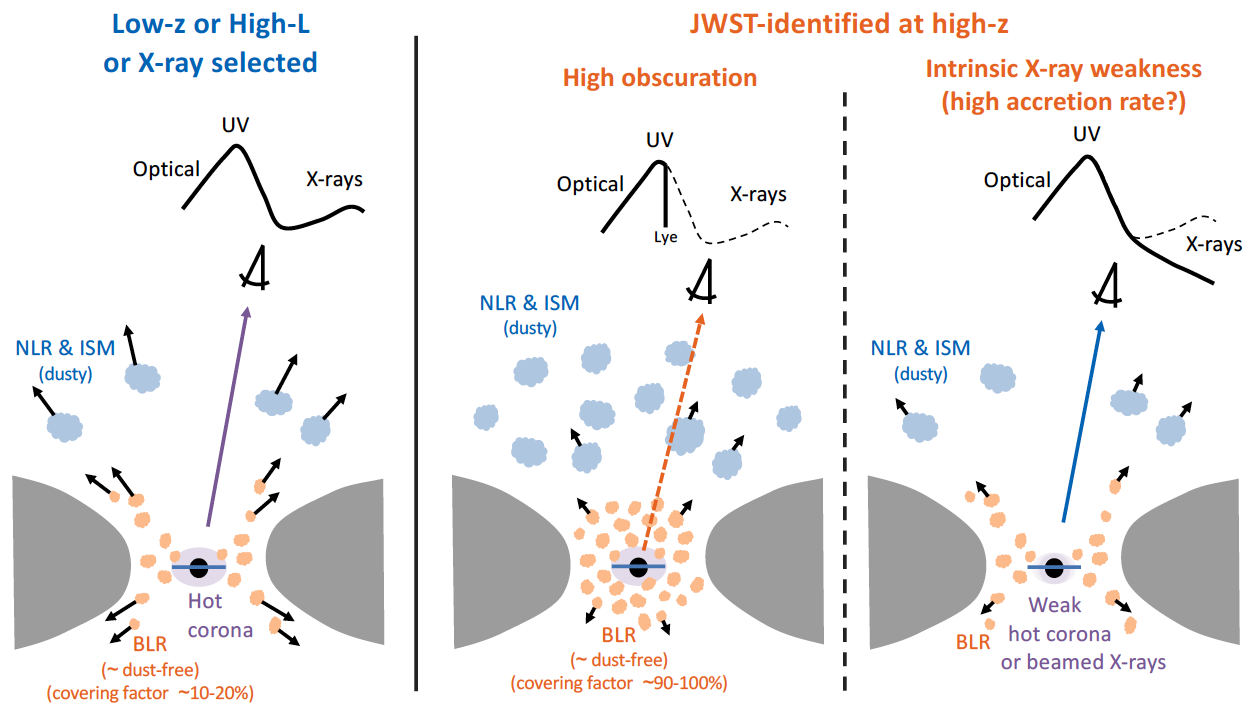}
	\caption{Sketch of possible scenarios illustrating the observed X-ray weakness of JWST-identified AGN at high-z as due to X-ray absorption from Compton thick clouds with low dust content (possibly in the BLR, centre) and intrinsic X-ray weakness (right).} 
	\label{fig:sketch}
\end{figure*}

\subsection{Compton thick, high covering, dust-poor absorption}

At the redshifts probed by the many AGN in our sample (z$\sim$4-7), the most sensitive Chandra energy range (0.5-2~keV) probes the 4-14~keV rest-frame range, and even higher energies for the most distant AGN in our sample. 
If the lack of detections is due to X-ray absorption, then this would imply Compton thickness, i.e. $N_H>10^{24}~cm^{-2}$. If the column density is high enough to totally absorb the direct X-ray spectrum, then the observed X-ray spectrum depends on the amount of cold and warm scattered emission. The latter depends strongly on the covering factor and on the column density of the reflecting medium (and on whether the reflected light is absorbed too) which, in the 2-10~keV band, can go from a few percent of the direct radiation, to potentially zero.

In the local Universe the fraction of obscured AGN is about 80\% and, of these, the fraction of Compton thick AGN is estimated to be of the order of 50\% \citep{Risaliti2002,Maiolino1998,Bassani1999,Ananna2019}.
There have been studies suggesting an increase of the obscured fraction at high redshift, and also an increasing fraction of Compton thick AGN, especially in the low-luminosity regime \citep{Buchner2015,Ananna2019,Gilli2022,Peca2023X, Signorini23}.

The potential issue with the JWST findings is that, in order to entirely explain the X-ray weakness with the Compton thick scenario:
\begin{enumerate}
    \item most of the newly discovered AGN should be Compton thick, i.e. the Compton thick material should be covering nearly 4$\pi$ the AGN;
    \item given that obscuration affects most type 1 AGN too, the Compton thick material should be dust-poor, or even dust-free.
\end{enumerate}

To give a more quantitative assessment of the latter issue, for a standard gas-to-dust ratio and Galactic extinction curve, a Compton thick medium ($N_H>10^{24}~cm^{-2}$) should give $A_V>500$ \citep{Maiolino2001a}. Even taking into consideration that at high redshift these galaxies are characterised by metallicities of the order of 0.1 solar, and that the dust-to-metal ratio is lower by a factor of about $\sim 2$ at such low metallicities \citep{DEugenio2023,Tacchella2024}, Compton thick absorption should still result in $A_V>25$. Although some dust reddening is typically seen in most high-z type 1 AGN found by JWST, these are in the range of $A_V \sim 0.5-4$ \citep{Maiolino23c,Matthee2023,Harikane_AGN,Ubler2023,Juodzbalis2024}, and anyway $A_V>25$ would certainly totally absorb the broad lines.

A major mismatch between X-ray absorbing column density and dust extinction in AGN, up to a factor of 100 relative to what is expected from the Galactic dust-to-gas ratio, has been known for a long time \citep{Maiolino2001a}. Although peculiar dust properties in the dense nuclear region of AGN may be partially responsible for this effect \citep{Maiolino2001b,Gaskell2004}, extensive X-ray variability studies have clearly demonstrated that the bulk of the X-ray absorption, especially at high column densities, happens within the dust sublimation radius, on scales typical of the BLR \citep{Risaliti2002,Risaliti2005,Risaliti2011,Maiolino2010,Nardini2011}.
More precisely, when probed with high enough temporal resolution, and with high enough signal-to-noise ratio, the observed X-ray absorption is seen to be highly variable, and even changing from Compton thick to Compton thin, consistent with being associated with clouds of the BLR transiting along the line of sight. The BLR is inside the sublimation radius, hence dust-free. On the other hand, both the X-ray variability studies, and photoionization modelling, indicate that the BLR clouds have large columns of gas, generally in excess of $10^{24}~cm^{-2}$. Therefore, absorption by the BLR clouds along our line of sight can very naturally explain extremely high values of $N_H/A_V$, and this may be the case also for the AGN newly discovered at high redshift by JWST. The latter do actually show some degree of reddening, so it is possible that some of the absorption is also associated with gas outside the BLR and in the host galaxy.

In summary, absorption by BLR clouds, and possibly additional absorption in the host galaxy, can account for extremely Compton thick X-ray obscuration with modest (or no) dust extinction in the optical/UV. However, the issue with the JWST-discovered AGN at high-z is that, if this is the explanation for their X-ray weakness, given the large number of AGN without X-ray detection, the covering factor of the BLR clouds should be close to unity. This would be a quite different BLR distribution relative to the local AGN or high luminosity quasars at high-z, as in these cases the BLR covering factor is estimated to be of the order of 10\%--20\% \citep{Netzer1990,Peterson2006}, although some recent works model the BLR with a covering of up to 50\% \citep{Guo2020,Ferland2020}, possibly as a function of the accretion rate (this aspect will be discussed later in this section). Additionally, the BLR in local and high-luminosity AGN is found to be primarily distributed in a disc-like configuration, probably co-planar with the obscuring torus (or disc of the host galaxy), hence making it much more unlikely for the line of sight to cross a BLR cloud without being totally obscured also by dust in the torus
\citep{Mannucci1992,Elvis2000,Maiolino2001c,Gravity2018,Pancoast2018}.

Therefore, a possible scenario is that, unlike local AGN and high luminosity quasars, in the high-z AGN newly discovered by JWST the ($\sim$ dust-free and Compton thick) BLR clouds are distributed more isotropically around the accretion disc and with a much larger covering factor, possibly approaching unity. Fig.~\ref{fig:sketch} schematically illustrates this scenario and in the following subsection we further investigate its plausibility.

Finally, we shall mention that according to \cite{King2025} a highly Compton thick envelope of the accreting black hole is expected in a scenario of super-Eddington accretion. This will be discussed further in section \ref{sec:NLSy1_superEdd}.

\begin{figure}
    \includegraphics[width=1.0\linewidth]{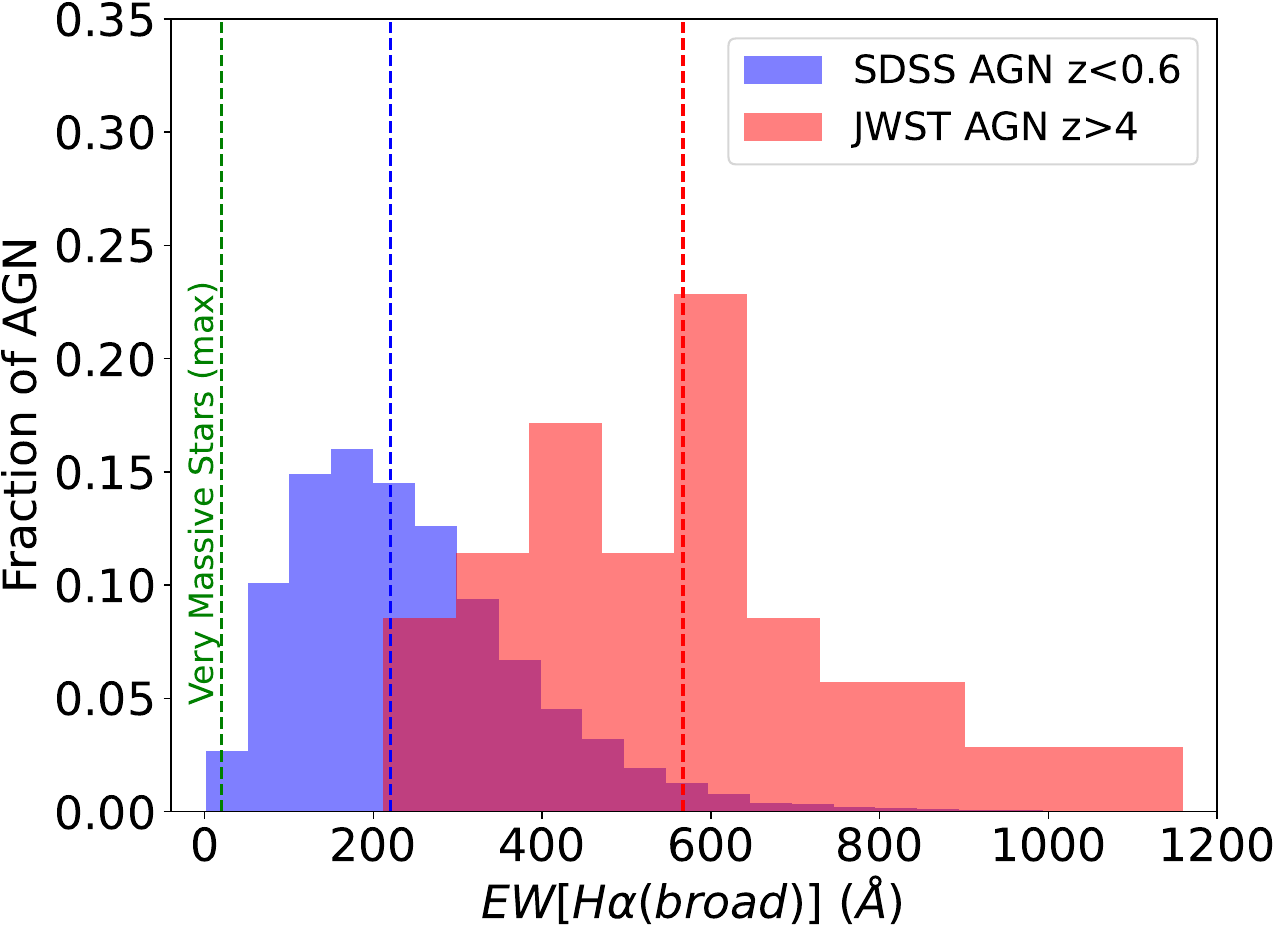}
	\caption{Distribution of the (rest-frame) equivalent widths of the broad component of H$\alpha$ for the AGN identified by JWST at high redshift (red) compared with AGN and quasars at low redshift (blue). The vertical red and blue dashed lines show the median of the two distributions. The vertical green dashed line indicate the maximum equivalent width expected from Very Massive Stars.} 
	\label{fig:EW_Ha}
\end{figure}

\subsubsection{High equivalent width of the H$\alpha$ broad component}

The BLR clouds absorb the ionizing continuum from the accretion disc and re-emit such radiation in the recombination nebular lines and continuum. In particular, the H$\alpha$ luminosity is proportional to the luminosity of the ionizing photons, modulo the covering factor. This means that, if the intrinsic SED of AGN (from optical to UV ionizing bump) is the same, then the equivalent width of the broad component of H$\alpha$ should be a tracer of the covering factor of the BLR clouds.

Fig.~\ref{fig:EW_Ha} shows the distribution of the $EW(H\alpha _{broad})$ for the type 1 AGN newly discovered at high-z by JWST (red histogram), compared with the same distribution for low redshift AGN and quasars from SDSS \citep[blue histogram, ][]{Lusso2020}. Clearly, the distribution of $EW(H\alpha _{broad})$ of the high-z AGN found by JWST is shifted to much higher values relative to the low-redshift AGN. One should also take into consideration that in many of the AGN found by JWST the optical continuum is contributed, and often dominated, by the stellar light from the host galaxy \citep{Maiolino24_GN-z11,Maiolino23c,Juodzbalis2024}. Therefore, the EW estimated for the JWST galaxies are actually lower limits. 

The vertical dashed lines in Fig.~\ref{fig:EW_Ha} indicate the medians of the two EW distributions, which are 200\AA\ for low-z AGN and quasars, while it is 570\AA\ for the high-z AGN identified by JWST. 
As mentioned above, the observed EWs are likely lower limits of the intrinsic EW, without galactic contamination. For this reason,
the ratio between the two distributions is at least 2.6. If the covering factor of low-z, optically/UV selected AGN was $\sim$40\%, then this ratio would imply that the BLR in the JWST-selected AGN would reach a covering factor of $\sim$100\%. However, as discussed above, the covering factor of the BLR in low-z, optically/UV selected AGN is uncertain, with estimates ranging from 10\% to 50\%. So we can only quantitatively say that the covering factor of the JWST-selected AGN is at least a factor of 2.6 higher, on average.

\begin{figure}
    \includegraphics[width=1.0\linewidth]{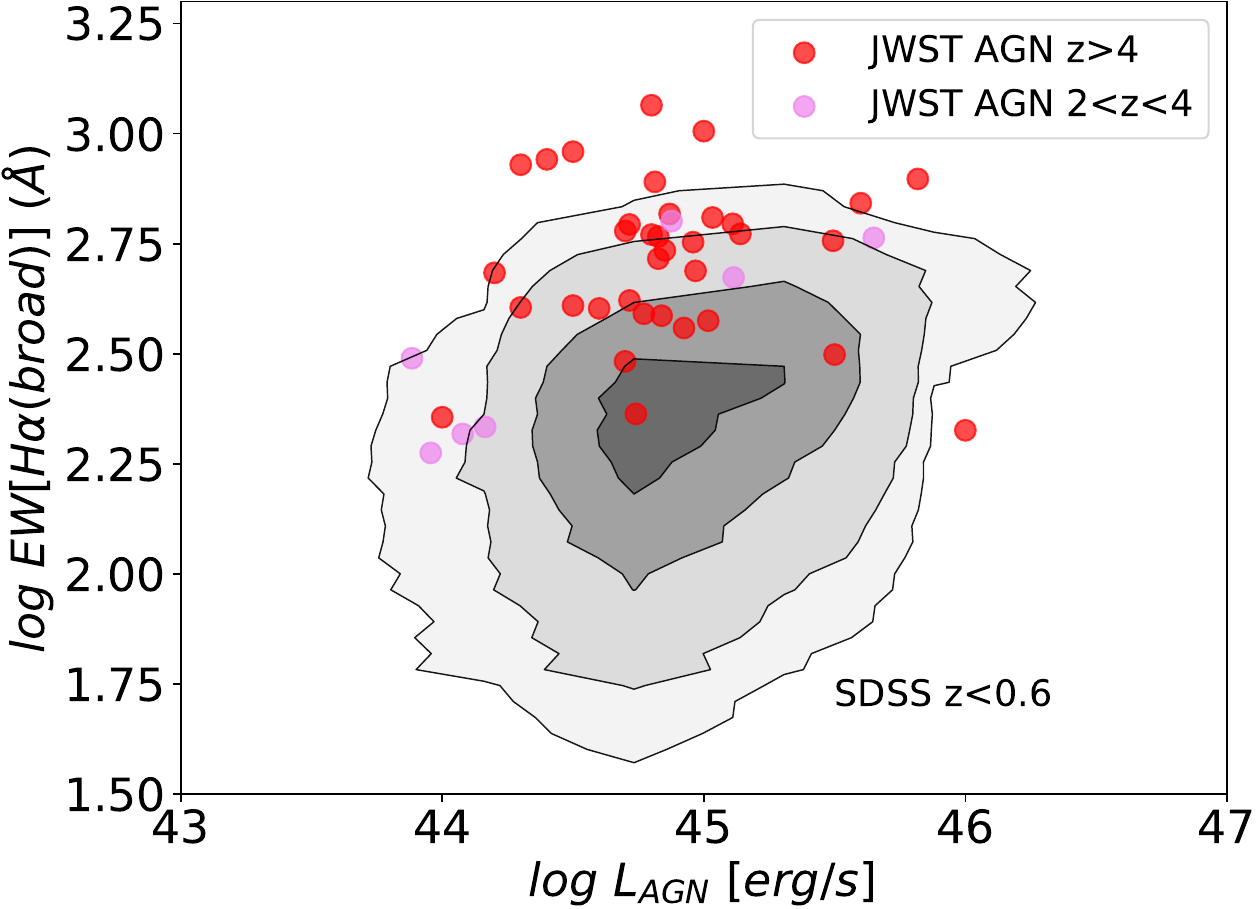}
	\caption{Equivalent width (rest-frame) of the broad component of H$\alpha$ versus AGN bolometric luminosity, for the AGN identified by JWST at high redshift (red for those at z$>$4, violet for those at 2$<$z$<$4), compared with the distribution AGN and quasars at low redshift (gray contours).} 
	\label{fig:EW_Ha_Lbol}
\end{figure}

As $EW(H\alpha _{broad})$ has some dependence on  AGN luminosity, in Fig.~\ref{fig:EW_Ha_Lbol} we show the equivalent width distribution as a function of $L_{bol}$. Even when taking into account the dependence on $L_{bol}$, the AGN found by JWST are significantly offset relative to standard AGN.

We have also tested the potential dependence on the accretion rate, in terms of $L/L_{Edd}$. This is shown in Appendix \ref{app:ew_acc_rate}. Also in this case, the EW(H$\alpha_{broad}$) is found to be larger than the distribution of optically selected AGN.

\begin{figure*}
    \includegraphics[width=0.85\linewidth]{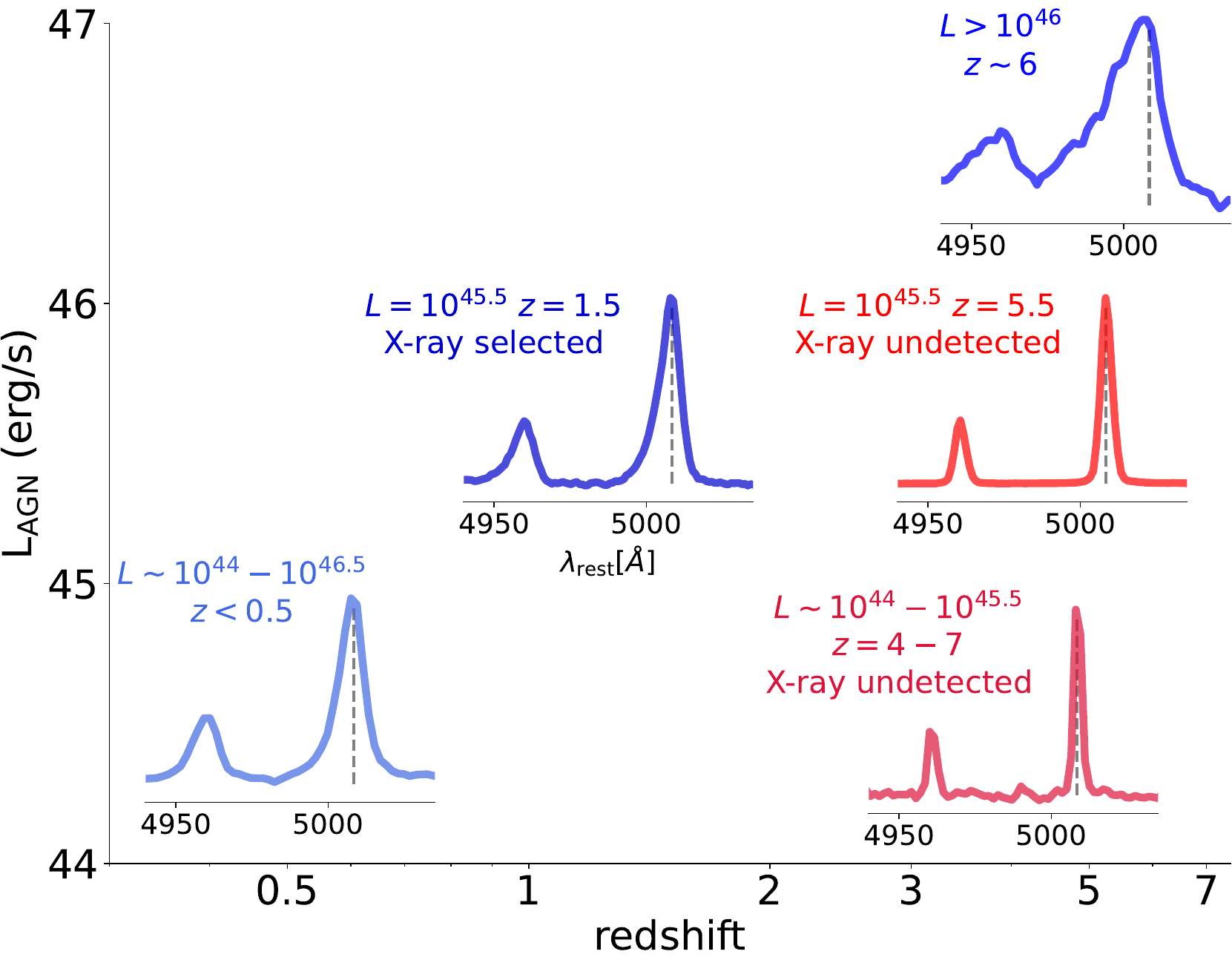}
	\caption{Comparison between the profiles of the [OIII]5007 doublet for different classes of galaxies. The blue lines indicate pre-JWST selected objects, specifically:  
    stack of low redshift AGN with intermediate luminosities, the X-ray selected AGN XID2028 at z=1.5, and stack of luminous (optically selected) quasars at z$>$6. The red lines show some of the JWST-identified AGN:
    stack of the (X-ray undetected) JADES type 1 sample in \citet{Maiolino23c} at $4<z<7$; 
    spectrum of the type 1.8 GS\_3073, 
    X-ray undetected, AGN at z=5.5. The high-luminosity, X-ray selected and low-z AGN, have broader [OIII]
    profiles and all have a prominent blue wing, typical signature of ionized outflows, while the AGN discovered by JWST at high-z have a narrow and symmetric profile, indicating much less prominent outflows, if any.
 } 
	\label{fig:comp_oiii}
\end{figure*}

\subsubsection{Additional indications of large gas columns and large absorbing densities}

Out of the four X-ray detected objects in our sample, the two detections that are deviating from the local $k_{bol,X} - L_{Bol}$ relation, i.e. XID403 at z=4.76 
GN 721 at z=2.94,
 have X-ray spectra consistent with being Compton thick.
%\citep[the latter was spectroscopically identified by ][ but the type 1 identification was obtained with JWST]{Gilli2011}.
Of the other three detections,
GS 49729 (z=3.2) and GS~209777 are not absorbed, but fully consistent with the local $k_{bol,X} - L_{Bol}$ relation,
while GS 21150 (included in the JADES sample because X-ray selected) is too faint to allow an estimate of the column density.

Outside the GOODS fields, the only X-ray AGN spectroscopically identified by JWST is UHZ-1, which is heavily Compton thick with $N_H\sim 10^{25}~cm^{-2}$  \citep{Bogdan23}.

\cite{Kocevski_AGN} identified two additional AGN that were X-ray selected: PRIMER-COS 3982 at z = 4.66, which was however previously spectroscopically identified and confirmed from ground, and JADES 21925, which has a photometric redshift of z = 3.1, but not spectroscopically identified. In these two cases the inferred column densities are substantial, $N_H\sim 10^{23}~cm^{-2}$, although not Compton thick.

Taking into account that X-ray detections are biased against heavily obscured systems, it is remarkable that the two high-z X-ray AGN identified by JWST, for which spectral analysis was possible, are Compton thick. These might be the tip of the iceberg of a much larger population of Compton thick AGN.

It is additionally interesting, among the JWST-discovered AGN, the emergence of type 1 AGN that show absorption features in the profiles of the broad H$\alpha$ and/or H$\beta$. Absorption of the Balmer lines is extremely rare at low redshift, less than about 0.1\% of the AGN population  
\citep{Shi2016, Zhang2015,Schulze2018,Williams2017}. 
However, the JWST spectra are revealing an increasing number of these Balmer line absorption features in type 1 AGN. The AGN with the lower limit on $L_{BOL}/L_X$ ($>10^4$, Fig.\ref{fig:Rbol_type1}) has prominent absorption of H$\alpha$ and H$\beta$ \citep{Juodzbalis2024Rosettta}. \cite{Matthee2023} found two cases of H$\alpha$ absorption in their sample of 20 broad line AGN, i.e. 10\% of their sample. \cite{Kocevski2024} find that 22\% of their broad line AGN, hosted in Little Red Dots, have some H$\alpha$ in absorption. In the JADES spectroscopic data release, there are three type 1 AGN with H$\alpha$ (and H$\beta$ in one case) in absorption, out of a sample of about 30 type 1 AGN
\citetext{Juod\v{z}balis in prep., \citealp{DEugenio2024}}. It should be noted that these fractions are lower limits, as detecting H$\alpha$ or H$\beta$ in absorption requires high S/N and, at least, medium resolution spectroscopy (R$\sim$1000) to disentangle the narrow and broad line emission profile from absorption. Those objects characterised by Balmer absorption are marked with magenta open circles in Fig.~\ref{fig:Rbol_type1}  and are indeed all extremely X-ray weak.
As n=2 is not a metastable level, observing H$\alpha$ or H$\beta$ in absorption  requires, in addition to temperatures of $T\sim 1\text{--}2\times 10^4$~K, very large gas densities of n$>10^9~cm^{-3}$, typical of the BLR \citetext{\citealp{Williams2017,Juodzbalis2024Rosettta}}. Therefore, the unexpectedly high fraction of type 1 AGN with H$\alpha$ or H$\beta$ in absorption indicates that at least in 10\% of the cases we are likely seeing the optical continuum through BLR clouds, hence through a column of gas that is likely Compton thick. Note that detecting Balmer absorption is an extreme case that likely implies Compton thickness, while the opposite is not true: Compton thick absorption does not necessarily imply Balmer absorption.

 While the fraction of broad-line, type 1 AGN in the local/low-z Universe with Balmer absorption is very low, it is intriguing to note that the prototypical broad-line Seyfert 1 galaxy, NGC~4151, does show prominent Balmer absorption \citep{Hutchings2002_NGC4151} and also strong (and variable) X-ray absorption, with absorbing column densities approaching $10^{24}~cm^{-2}$ at times, resulting in significant X-ray weakness \citep{Zoghbi2019_NGC4151}. The location of NGC~4151 in Fig.\ref{fig:Rbol_type1} is shown with a light green diamond.

We finally note that unobscured lines of sight passing through the BLR clouds should also result in absorption of the UV resonant lines, such as CIV, SiIV and AlIII.
GN-z11 is one of these cases \citep{Maiolino24_GN-z11}, but there are very few other cases. However, most of the type 1 AGN identified by JWST
show some degree of optical reddening ($\rm A_V\sim 0.5-4$), which implies that, for most of them, the intrinsic UV radiation from the accretion disc is mostly absorbed. Therefore, the observed UV radiation is either dominated by stars in the host galaxies or scattered radiation, hence it would not be possible to see the UV absorption features associated with dense gas along the line of sight. Additionally, in order to be detected, the absorption features have to be shifted relative to the host galaxy rest frame, or else they would be filled in by the emission component of the same transition. If, as discussed, clouds linger in the nuclear region and do not have significant velocities, then they would be hard to see even at medium/high resolution spectroscopy, and impossible to see with prism spectroscopy. Finally, one should take into account that the blue part of the NIRSpec spectra has the lowest sensitivity and, typically, the noisiest. A number of potential absorption features are often seen, but difficult to validate with confidence.

\subsubsection{Low AGN ejective feedback}

One question is why the covering factor by dense gas (possibly from the BLR) should be much larger in the intermediate-luminosity AGN at high redshift discovered by JWST.
One possibility is that the ejective feedback in these intermediate-luminosity AGN at high-z is much reduced, hence dense clouds linger in the vicinity of black holes. One indication going in this direction is that the population of AGN at high-z discovered by JWST seems to generally lack the prominent broad, blueshifted wings of [OIII] tracing ionized outflows and typically seen in the vast majority of AGN at low-z, or at high-z, but with high luminosity or X-ray selected \citep{Brusa2015,Carniani2015,Cresci2015,
Bischetti2017,Leung2019,Kakkad2020}.

We illustrate this property more quantitatively in Fig.~\ref{fig:comp_oiii}, where we overplot the normalized and continuum subtracted [OIII] profiles of different classes of AGN. The blue profiles show the [OIII] doublet for AGN identified and selected before JWST. Specifically:  the bottom-left plot shows the stack of AGN and quasars at z$<$0.6 from SDSS \citep{Lusso2020} (R$\sim$2000), in a luminosity range $10^{44}<L_{AGN}<10^{45.5}~erg/s$, hence matching the type 1 luminosity range found with JWST by \cite{Maiolino23c} (details on the stacking procedure are given in Appendix \ref{app:stack}); the top-right plot shows plot shows the stack of the luminous SDSS quasars at z$\sim$6 observed with JWST by \cite{Marshall2023}, \cite{Eilers2023} and \cite{Loiacono2024} (R$\sim$2000--2700); the central plot shows the spectrum of XID2028, an X-ray selected luminous AGN at z=1.5, observed with JWST by \cite{Cresci2023} and \cite{Veilleux2023} (R$\sim$2700).
The red lines show [OIII] profiles of JWST-identified AGN (all X-ray undetected). Specifically: the bottom-right plot shows the stack of the high resolution spectra (R$\sim$2700) of the type 1 AGN in JADES presented in \cite{Maiolino23c} (R$\sim$2700, note that high resolution spectra are often unavailable for other NIRSpec MSA surveys, which often use only R=1000 or even Prism); the central-line plot shows the spectrum of GS\_3073, the X-ray undetected type 1.8 AGN identified by \cite{Ubler2023} (R$\sim$2700) and previously discussed. Clearly, the AGN selected pre-JWST, at low-z, or high-z but high luminosity, or X-ray selected, have a broader [OIII] profile and, most importantly, a prominent [OIII] blue wing, tracing high velocity ionized outflows. On the contrary, the X-ray undetected, JWST-identified AGN have narrower [OIII] profiles and, most importantly, they lack the prominent blueshifted wings seen in other AGN. It should be noted that GS\_3073 does have some very faint wings, actually redshifted, which in the IFS cube have been interpreted as an ionized weak outflow \citep{Ubler2023}; however, in terms of the ionized gas mass relative to the narrow component, it is far less prominent than in other AGN.  The same applies to the stacked JWST type 1 spectrum: a very detailed and in depth analysis does reveal a very faint wing of [OIII] \citep{Trefoloni2024Fe}, but far weaker than seen in the stack of their low redshift counterpart.
Overall, the profile of the [OIII] line suggests that the high-z AGN identified by JWST are driving very weak ionized winds, if any.

The large scale ionized winds that are found in other AGN are thought to arise both from direct radiation pressure on dusty clouds and by energy driven winds originating from the central region and involving the BLR \citep{Fabian2012,King_Pounds_2015,Costa2018,Costa2020}. The lack of ionized winds on large scales inferred from the [OIII] symmetric profile suggests that also the BLR is characterised by weak winds. Directly identifying and characterizing winds of the BLR, or of gas distributed on similar scales, is more difficult. Typically, these are traced via the resonant UV lines, such as CIV. There is no evidence of CIV blueshifted absorption in any of the AGN analysed in our sample; yet, this may not be meaningful, as the UV light is generally dominated by the stellar light in the host galaxy (and see also the NIRSpec sensitivity issues in the bluest band discussed in the previous section). The only exception is GN-z11 at z=10.6, which does show a clear CIV blueshifted absorption. The velocity ($\sim 800-1000 ~km/s$) indicates that it is clearly driven by the AGN. However, it is far slower and shallower than typically seen in AGN and quasars with similar accretion rates \citep{Bischetti2022,Gibson2009,Maiolino2004}, indicating that even in this case the outflow is much milder than in other AGN. 
GS\_3073 may also have a mild nuclear outflow traced by NV, but very weak \citep{Ji2024}.

The H$\alpha$ absorption discussed above, seen with JWST in a few type 1 AGN of our sample, and most likely associated with absorption by BLR clouds along the line of sight, is blueshifted only by a few 100 km/s in three cases, redshifted in one case, and at rest frame in another case \citetext{\citealp{Matthee2023}, Juod\v{z}balis in prep., \citealp{DEugenio2024},\citealp{Kocevski2024}}. Although these may not be representative of the entire sample, these direct tracers of the BLR kinematics, indicate that in these objects the BLR is either outflowing with very low velocities, stalling, or even inflowing.

In summary, the various lines of evidence support a scenario in which the ejective feedback in these JWST-selected AGN at high-z (which are the majority at high redshift) is less effective than for other classes of AGN studied in the past.

The reason why feedback is reduced in these low/intermediate luminosity AGN at high-z should be explored in detail in dedicated studies. We only speculate on the possibility that this could be due to the low metallicity characterising these AGN at high-z. Wherever the gas clouds are dusty, lower metallicity means lower dust content and, therefore, lower radiation pressure on dust. This should not affect the clouds in the BLR, which are dust-free. However, a reduced content of metals should also reduce the winds in the nuclear region, as these are primarily driven by line locking on the metal ionized species.
At lower redshift, the higher metal enrichment and dust content are expected to foster outflows. Similarly, high-luminosity quasars probably manage to develop outflows even at high redshift, both because the radiation pressure is anyway very high and because they are hosted in more massive systems that have been enriched more rapidly, particularly in the nuclear region \citep{Juarez09,Ji2024,Costa2018,Costa2020}.

Since these high-z galaxies are extremely compact and characterized by high gas densities \citep{Tacchella2023,Ji2024}, it could also be that the nuclear outflows struggle to develop to large scales and remain confined to the very central (pc-scale) region.

We note that ionized outflows have been detected in intermediate luminosity AGN at z$\sim$2 \citep[e.g.][]{Genzel2014,Forster2014} and neutral
outflows have been identified in JWST-discovered AGN at z$\sim$2--3
\citep[primarily via sodium absorption][]{DEugenio2023,Belli2023,Davies2024}. However, these are massive ($M_{star}>10^{10}~M_\odot$), metal enriched galaxies, quite different from the bulk of the AGN population discovered by JWST, which are hosted in low mass ($M_{star}\sim 10^{8-10}~M_\odot$) and metal poor host galaxies. Therefore, these are probably two different populations of AGN.

\subsection{Intrinsically weak X-ray emission}

Another possible factor contributing to the observed weakness of the X-ray emission can be the intrinsic weakness of the AGN emission. In this subsection we investigate a few possible scenarios.

\subsubsection{Narrow Line Seyfert 1 and high accretion rate}
\label{sec:NLSy1_superEdd}

Narrow Line Seyfert 1 (NLSy1) are defined as type 1 AGN whose `broad' permitted lines have widths that are significantly narrower ($FWHM\sim 500-2000~km/s$) than classic type 1 AGN and quasars, but still broader than the forbidden lines in the same objects (e.g. [OIII]) \citep{osterbrock+pogge1985,veron-cetty+2001,Berton2016}. This class of AGN is thought to be characterized by high accretion rates, close to the Eddington limit, or even super-Eddington \citep{Mathur2000,Collins2004}. NLSy1
are well known to have a steeper X-ray spectrum than normal AGN, and this results in an X-ray weakness in the hard 2-10~keV band, by up to a factor of $\sim$5 \citep{Vasudevan07,Tortosa2022,Tortosa2023}. Their
softer spectrum is consistent with theoretical expectations that highly accreting AGN should have a softer ionizing spectrum. Some of the local NLSy1 are shown on the $L_{bol}/L_X$ vs $L_{bol}$ diagram in Fig.~\ref{fig:Rbol_type1} with purple diamonds, illustrating that some of them are indeed quite weaker than standard AGN in the 2--10~keV band.

\begin{figure}
    \includegraphics[width=\linewidth]{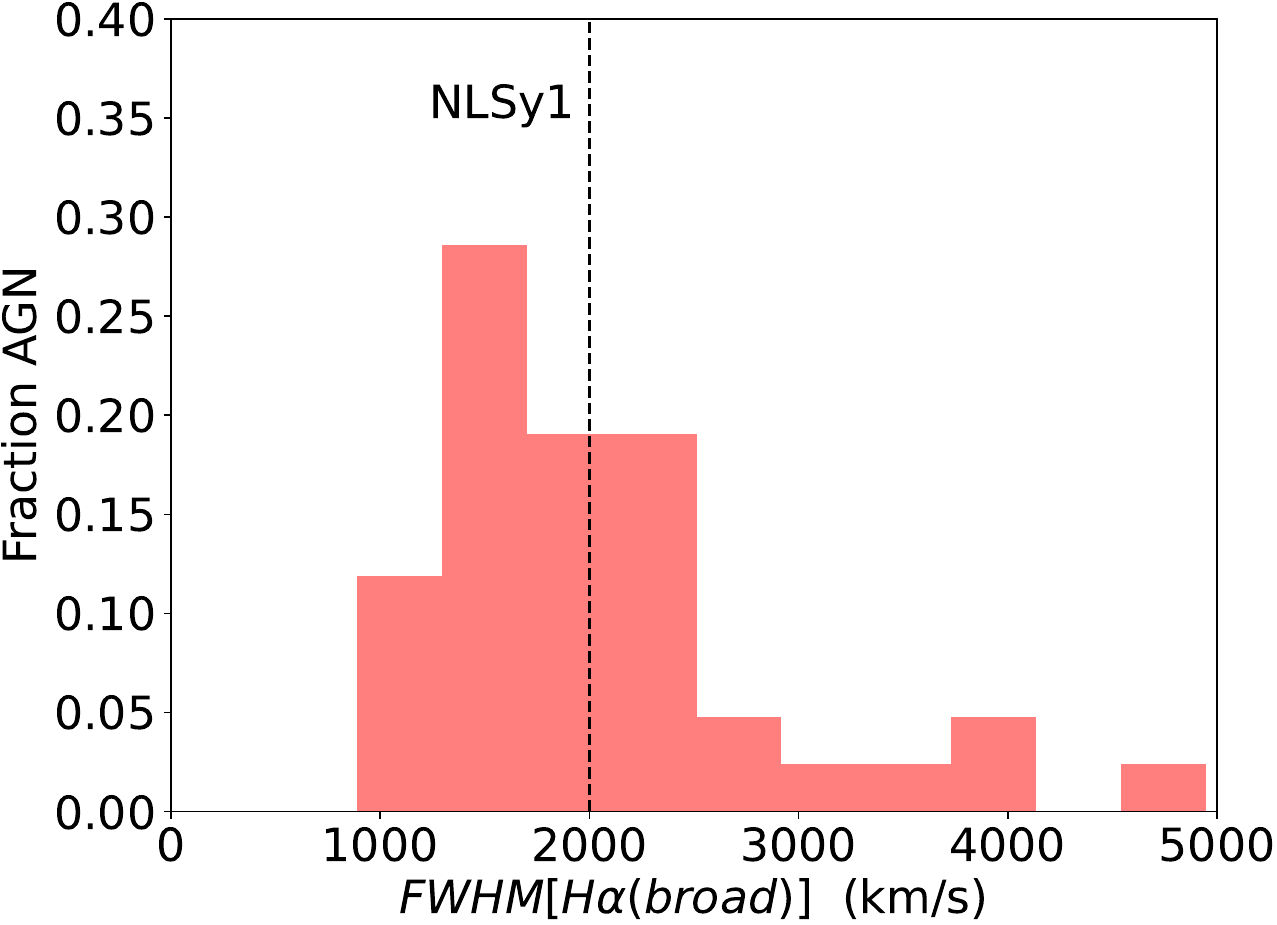}
	\caption{Distribution of the widths of the broad component of H$\alpha$ among the type 1 AGN identified by JWST at high redshift. The vertical dashed line indicates the limit below which AGN are classified as Narrow Line Seyfert 1.} 
	\label{fig:FWHM}
\end{figure}

Most of the AGN newly discovered by JWST are characterised by relatively narrow `broad' permitted lines, and most of them are formally in the range of NLSy1. This is shown in Fig.~\ref{fig:FWHM}, which provides the broad H$\alpha$ FHWM distribution of the JWST-identified AGN at high-z, and indicates that many of them should be classified as NLSy1.

Additionally, a good fraction of the JWST-selected AGN have fairly high accretion rates, close to or even above the Eddington limit \citep{Maiolino23c}. At low redshift it has been shown that highly accreting black holes can be X-ray weak \citep{Laurenti2022}.  Theoretically, it has been recently proposed that highly accreting black holes, possibly above the Eddington limit, should have their X-ray emission beamed in a narrow angle, hence X-ray weak along most of the lines of sight
\citep{Pacucci2024_Xrays,Madau2024Xray,King2025}.

Therefore, it is very likely that part of the reasons why the JWST discovered AGN at high redshift are so X-ray weak may be associated with intrinsic weakness, as for local NLSy1 and highly accreting AGN.

Within this context, it is interesting to note that X-ray observations of high-z quasars do show evidence for a steeper spectrum, similar to local NLSy1 \citep{Vito2019,Zappacosta2023,Wolf2023}. This may suggest that a steep X-ray spectrum may be a common feature of AGN in the early Universe and that this may make them more difficult to detect them at high redshift (where we probe higher energies rest-frame), unless the AGN has high, quasar-like luminosities.

However, as illustrated in Fig.~\ref{fig:Rbol_type1}, the similarity with NLSy1 cannot fully account for the X-ray weakness, especially the extremely high $L_{bol}/L_X$ seen in type 1 stacks and some individual cases. Additionally, there are examples of type 1 AGN identified by JWST that are clearly accreting well below the Eddington-limit and that are X-ray undetected, with $k_{\rm bol,X}$ more than one order of magnitude higher than the standard relations. The dormant black hole GN-1146115 at z=6.67 \citep{Juodzbalis2024}, marked with a black solid circle in Fig.~\ref{fig:Rbol_type1}, is one of these cases.

Yet, although an X-ray soft, steep spectrum cannot fully account for the observed X-ray weakness, it can certainly contribute to making the observed X-ray emission undetected (Fig.\ref{fig:sketch}),
especially at high redshift, where we are probing the hard portion of the X-ray emission, 
possibly even with lower absorbing column, and relaxing the requirement of Compton thickness.

\subsubsection{Reduced hot corona}

Most models ascribe the production of hard X-rays to the inverse Compton scattering of the UV photons by a hot corona above the accretion disc \citep{Haardt1991}. Therefore, one possibility is that these newly discovered AGN at high-z are lacking a hot corona, or that it is greatly reduced. Although intriguing, testing or even speculating on this possibility is difficult.

In the local Universe there have been cases of destroyed corona \citep{Ricci2020}, but it is not clear what caused the event and, in particular, in these cases there is nothing obviously connected to the high redshift Universe.

It is not clear how the hot corona is produced, but models expect that it should result from the magnetic field lifting material from the accretion disc. It could be that at high redshift, and around lower-mass black holes, the magnetic field has not yet developed properly. As far as we are aware, there are no models or simulations describing the corona of AGN at high redshift and, in particular, in intermediate luminosity AGN.
Yet, there is recent major and very promising progress in the development of zoom-in radiation-magnetohydrodynamic simulations, down to very small scales \citep[$\sim$300 Schwarzschild's radii][] {Hopkins2024a,Hopkins2024b,Koudmani2023,Martin-Alvarez2023}. However, none of these simulations is yet probing the magnetic field in the accretion disc for black holes at high redshift in the mass range probed by our study.

\subsection{Non-AGN contaminants}

As discussed previously, most of the X-ray undetected AGN discovered by JWST have overwhelming and unambiguous evidence for the presence of AGN. The most spectacular case is GS\_3073 that, in addition to the prominent broad permitted lines of the H Balmer series, HI and HeII, also shows evidence for an array of 40 nebular emission lines, some of which coming from highly ionised species, and even coronal lines \citep[NV, NIV, ArIV, FeVII, FeXIII, FeXIV, ][]{Vanzella2010,Grazian2020,Ubler2023,Ji2024}, which unambiguously identify it as a luminous AGN. Despite that, it is undetected in the deepest region of the Chandra 7Ms field. However, not all AGN identified by JWST have the same wealth of information. Yet, as discussed, the lack of X-ray emission is common to different 
AGN selection criteria (broad lines, narrow line diagnostics, mid-IR excess), and therefore it is unlikely that it arises from mis-classification issues of all these methods. The lack of detection in the stacks would imply that the vast majority in each of those categories has been misclassified. In particular, in the case of type 1 AGN (which have the most stringent stacked constraint), the lower limit on $L_{Bol}/L_X$ inferred from the stack, which is more than two orders of magnitude above the value expected from the standard AGN SED, would imply that more than 99\% of the objects have been misclassified as AGN.

However, in the following, we discuss the scenario in which a fraction of the AGN identified by JWST might be misclassified.

\begin{figure}
    \includegraphics[width=\linewidth]{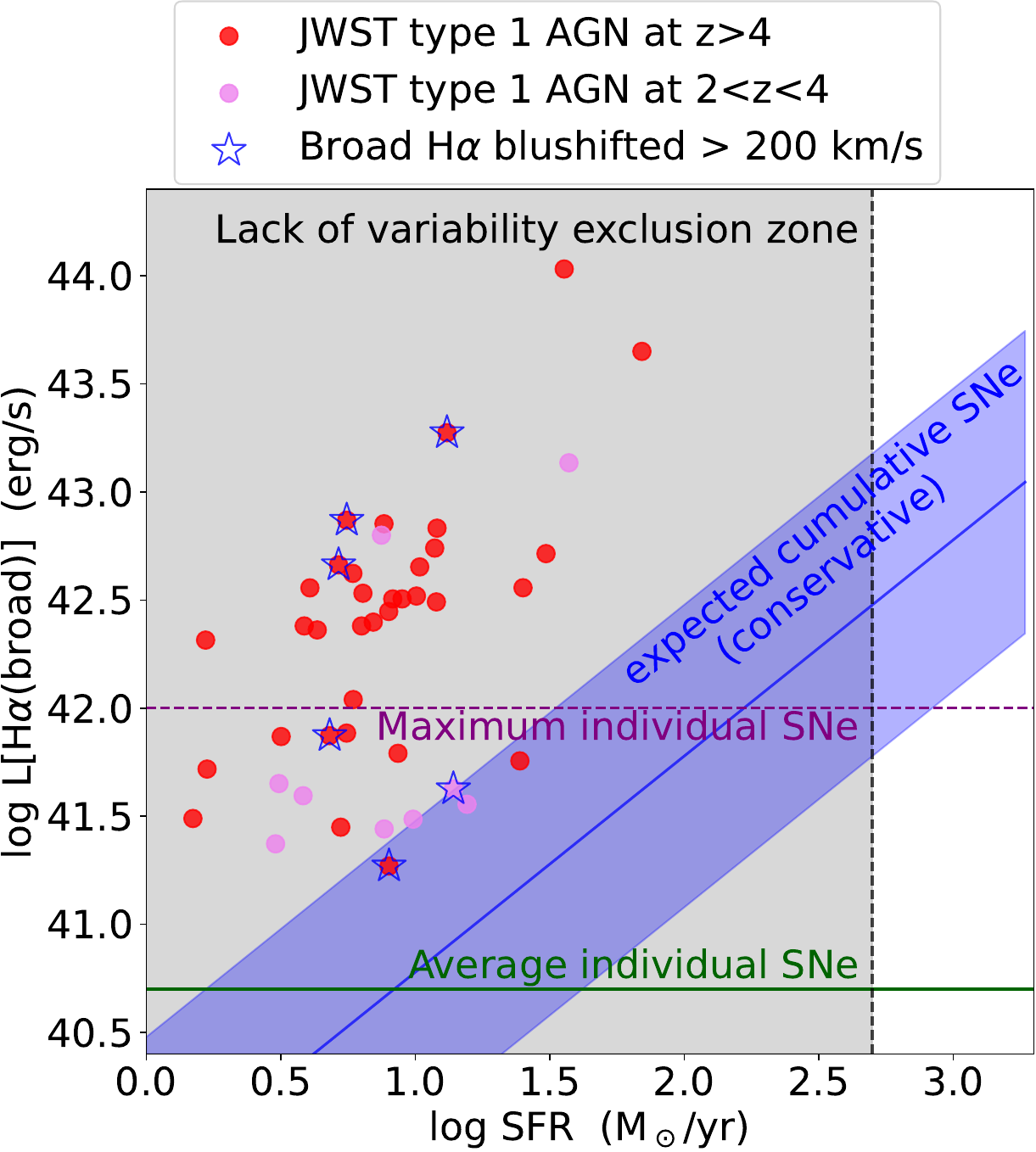}
	\caption{Constraints on the contribution core-collapse SNe to the broad component of H$\alpha$ in a diagram showing the luminosity of the latter as a function of the star formation rate (derived assuming the extreme case that the narrow H$\alpha$ is entirely due to star formation). Red and violet points indicate AGN identified by JWST at z$>$4 and at 2$<$z$<$4, respectively. The vertical dashed line and grey shaded region show the region in which contribution from multiple SNe is excluded by the lack of variability. The blue solid line and blue shaded region indicate the expected contribution to the broad component of H$\alpha$ from multiple supernovae in the case continuous star formation rate.
    The horizontal green line shows the average broad H$\alpha$ luminosity of core-collapse SNe, averaged over one year timescale, as inferred from the samples in \citet{Kokubo2019} and \citet{Taddia2013}. The horizontal purple dashed line shows the maximum luminosity of the broad component of H$\alpha$ seen in rare, superluminous SNe. The open starred symbols indicate objects whose broad H$\alpha$ is blueshifted by more than 200 km/s, as it is the case in the late phases of SN light curves.
	}\label{fig:SNe}
\end{figure}

\subsubsection{Core-collapse Supernovae}
\label{sec:sne}

Core-collapse Supernovae (cc-SNe) produce broad hydrogen emission lines which may potentially be mistaken for type 1 AGN. Yet, none of the type 1 AGN detected so far in the GOODS fields shows evidence for multi-epoch variability, indicating that individual SNe can be ruled out \citep[e.g.][DeCoursey et al. in prep.]{Maiolino23c,Juodzbalis2024,Ubler24}. One potential issue is that in some cases the first epochs come from HST, hence limited to wavelengths $< 1.8~\mu m$. Therefore, obscured SNe may not be identified by HST imaging. However, several of the AGN found in the GOODS fields do have multi-epoch observations with NIRCam (DeCoursey et al. in prep.), or by comparing IFS and NIRCam observations \citep{Ubler24}, and still do not show variability.
Anyway, we also note that, even if some degree of variability was found, this could also be ascribed to AGN variability. 

However, since the broad component of H$\alpha$ can be bright even for 2-3 years after explosion \citep{Kokubo2019,Gutierrez2017,Fransson2014,Taddia2013}, it is possible that in some of the targets with broad H$\alpha$ we are seeing the cumulative effect of multiple supernovae. In this scenario, the SNe would be so frequent that their (photometric, broad band) light curve would have to overlap in time to mimic the lack of variability. We have simulated this effect and, in order not to show variability at the level of 0.1 mag (sensitivity for variability), the galaxy should produce (in the photometric aperture) at least 10 SNe per year. This sets a lower limit on the SFR of $500~M_\odot/yr$. An upper limit on the SFR in the galaxies of our sample can be inferred by assuming the most conservative case that there is no AGN and that the observed H$\alpha$ corrected for extinction is entirely produced by star formation. Fig.~\ref{fig:SNe} shows the inferred distribution of (maximum) SFRs, together with the luminosity of the broad H$\alpha$; clearly, the maximum SFRs are well below the limit required by SN variability, indicated with a vertical dashed line (the region excluded by variability is indicated with gray shading). This comparison alone excludes significant contamination by core-collapse SNe, even in the scenario of cumulative, steady contribution.

Additionally, the solid blue line and blue shaded region in Fig.~ \ref{fig:SNe} indicate what would be the expected cumulative average luminosity of the broad component of H$\alpha$ as a function of SFR \citep[by conservatively taking an average SN broad H$\alpha$ luminosity of $10^{41}~erg/s$ and, conservatively, over a timescale of 3 years, see ][]{Kokubo2019,Taddia2013}, which is well below the value observed in the JWST type 1 AGN.

For completeness, we also show the maximum luminosity of the H$\alpha$ broad line seen, at peak, in some rare superluminous SNe \citep{Fransson2014,Gutierrez2017,Yan2015,Taddia2013,Pastorello2002,Kokubo2019}. We note however that the more typical peak luminosity is around $10^{41}~erg/s$. H$\alpha$ luminosities of $10^{42}~erg/s$ are much more rare and more short-lived than normal core-collapse SNe.

Finally, the SN interpretation is also problematic from the point of view of spectral properties. The broad component of H$\alpha$ in SNe is generally very broad, often exceeding 10,000~km/s
\citep{Kokubo2019,Nicholl2019,Pessi2023}, much broader than what is seen in most of the type 1 AGN identified by JWST (Fig.\ref{fig:FWHM}). In the early phases, H$\alpha$ is generally characterized by a P-Cygni profile \citep{Tartaglia2016,Gutierrez2017,Dickinson2024,Pastorello2002} and at later times is generally blueshifted \citep{Kokubo2019,Fransson2014,Ransome2021,Taddia2013,Gutierrez2017}, due to the formation of dust in the ejecta \citep{Schneider_Maiolino_review_2024} -- these properties are not observed in the type 1 AGN (although a few type 1 AGN in JADES do show blueshifted profiles, which will be discussed below). Additionally, the spectrum of core collapse SNe is often also characterised by other prominent broad features, such as MgI], CaII, [OI], FeII which are not seen in the JWST spectra \citep{Gutierrez2017,Nicholl2019,Nicholl2020,Pessi2023,Yan2015,Taddia2013}. We note that none of these features is seen even in the stacked spectra of type 1 AGN \citep{Trefoloni2024Fe}.

The only possibility to interpret some of the broad H$\alpha$ emissions as associated with SNe in some of the objects, is to assume the case of an individual superluminous SN that exploded several months before the epoch of the observation, and which has faded below the threshold for being detected as a variable source, but whose broad H$\alpha$ profile has remained visible on longer (few years) timescales, and detectable for longer period, as seen in some core-collapse (but not superluminous) SNe \citep{Taddia2013,Fransson2014,Kokubo2019}. This would require some more detailed modelling and fine tuning to identify the specific sub-class of SN and the timeframe in which the SN continuum variability would become undetectable (which is different from object to object) and the broad H$\alpha$ still detectable. This is beyond the scope of this paper and will be addressed in a separate, dedicated work. Here we only note that such cc-SNe for which the broad H$\alpha$ reaches the maximum luminosity ($\sim 10^{42}$ erg/s), hence most likely to be detected, are also those for which the broad H$\alpha$ is short lived and has a steeper light curve, hence most difficult to fit in the scenario above. In the very dense gaseous environment typical of high redshift galaxies, supernovae are expected to release their energy even more rapidly, hence their features should also be more short lived. Additionally, we note that in the scenario above, in which the H$\alpha$ broad is seen at late times, it should be seen significantly blueshifted \citep{Kokubo2019,Fransson2014,Ransome2021,Taddia2013,Gutierrez2017}, due to the formation of dust in the SN ejecta \citep{Schneider_Maiolino_review_2024}. However, most of the H$\alpha$ broad profiles are generally symmetric, possibly in some cases redshifted \citep{Harikane_AGN,Maiolino23c,Ubler2023,Ubler24,Juodzbalis2024,Matthee2023}. There are only very few cases in which the broad H$\alpha$ is blueshifted by more than 200 km/s, and these are marked with a blue star in Fig.~\ref{fig:SNe}. Out of these, three are well above the maximum H$\alpha$ luminosity at the peaks of rare, superluminous SNe. Three are below this maximum value. We cannot exclude that these three (out of $\sim$44) could be associated with the echo of superluminous SNe.

In summary, significant contamination by core collapse SNe to the population of type 1 AGN discovered by JWST is excluded on multiple grounds. However, a small fraction of the faintest type 1 AGN could be associated with the late echo of superluminous supernovae.

We finally note that superluminous SNe are also observed to emit X-ray with luminosities above our detection threshold in stack
\citep{Levan2013,Greiner2015}. This further confirms that they cannot account for the large population of type 1 AGN without X-ray emission.

\subsubsection{Very Massive Stars}

Young Very Massive Stars (VMSs, $M_*>100~M_\odot$) are also characterised by a broad component of the Balmer lines \citep{Martins2020,Martins2023}. 
However, in these cases the equivalent width of the broad component of H$\alpha$ is very small, typically a few \AA\ and always below 20~\AA \ \citep{Martins2020}, i.e. much smaller than what observed in the JWST sources (green vertical dashed line in Fig.~\ref{fig:EW_Ha}). It should be noted that the EW of the broad component of H$\alpha$ expected from Very Massive Stars embedded in unresolved stellar populations is actually much lower, due to the dilution from the other, lower mass stars.

\subsubsection{Ionized galactic outflows}

It has been suggested that, in the case of type 1 AGN, the broad component of H$\alpha$  is not tracing a BLR, and that it might be associated with fast outflows, with velocities of few thousands km/s, instead \citep{Yue2024}. However, in the case of SF-driven outflows the H$\alpha$ broad wings reach values of only a few hundred km/s \citep[e.g.][]{Genzel2011}, and not the thousands of km/s seen in our sample. More importantly, in the outflow scenario, such broad components should be seen even more prominently in the [OIII] line. On the contrary, the [OIII] profile does not show evidence for a broad component, which is not seen even in the stacked spectrum in Fig.~\ref{fig:comp_oiii}. This excludes the outflow scenario with high confidence.
The only possibility to suppress [OIII], to the level of not being seen relative to the prominent H$\alpha$ broad component, would be by assuming that the outflow has a metallicity below $10^{-2}~Z_\odot$. This would be an order of magnitude lower than the metallicity of the host galaxies of these AGN (typically $0.1~Z_\odot$) and would go in the opposite direction of what is observed in any other galactic outflows, whereby they are more metal enriched than the host galaxies that produce them.

\subsubsection{Tidal disruption events (TDE)}

This class of events would not really be `contaminants' as TDE are effectively AGN, simply the source of fuel is different, stripped stars rather than gas from the circumnuclear region. Anyway, it is unlikely that this class of phenomena can contribute significantly. To begin with, TDE are known to show strongly variable light curves \citep{Gezari2021,Lodato2011}, while, as mentioned in Sect~\ref{sec:sne}, there is no indication of variability. Secondly, more than 40\% of TDE emit strongly in X-rays, with luminosities $L_X>10^{42}~erg/s$ and up to $L_X\sim 10^{44}~erg/s$ \citep{Guolo2023}, hence above our upper limits in several AGN. 
Finally, the H$\alpha$ line profiles associated with TDEs \citep{Gezari2021} are typically significantly broader than observed in the JWST-identified type 1 AGN.
Yet, it is possible, as for the case of superluminous SNe, that in a fraction of objects we are witnessing a phase in which the continuum variation is no longer detectable, but a broad H$\alpha$  might still be visible in the spectra. This is however difficult to test as the statistics on the temporal evolution of the broad H$\alpha$ for TDE is less extensively studied than for SNe.

\subsubsection{Type 2 contaminants}

In the case of the type 2 AGN, none of the potential contaminants discussed above apply. However, since some of the classical diagnostics, such as the BPT, seem to break down at high redshift \citep{Harikane_AGN,Ubler2023,Maiolino23c}, searches for type 2 AGN have relied on alternative diagnostics. One of the most trustworthy is the detection of high ionization transitions, such as [NeV], [NeIV] and NV; however these are weak and detected only in a minority of objects \citep{Chisholm2024,Scholtz23}. The majority of the type 2 selected by \cite{Scholtz23} are identified either through a more restrictive version of the BPT diagrams (to minimize contamination by star forming galaxies with extreme conditions), or via diagnostic diagrams using UV transitions and based on photoionization modelling \citep{Feltre16,Nakajima22} \citep[see also ][]{Mingozzi2024}. While \cite{Scholtz23} are very conservative in their type 2 selection, contamination by star forming galaxies with extreme properties (high ionization parameter and harder ionizing spectra, for instance from WR stars) is still possible. However, since the stacked value of the $L_{Bol}/L_X$ for the type 2 sample is two orders of magnitude higher than expected from a standard AGN SED \citep[and even larger if assuming the bolometric correction provided by][]{Netzer19}, this would imply that more than 99\% of the type 2 AGN have been misidentified. This is quite extreme, even because a number of them are confirmed by multiple diagnostics and some of them with very high ionization lines.

In summary, although a contamination of the type 2 AGN sample from extreme star forming galaxies is possible, it is very unlikely that this can entirely explain the extreme X-ray faintness.

\section{Implications}

In this Section we discuss the implication of the X-ray weakness of the JWST detected AGN, as well as the scenarios proposed, for our understanding of AGN properties and their evolution with cosmic time.

\subsection{X-ray background}

\cite{Padmanabhan2023} had pointed out that {\it if} the large population of AGN newly discovered by JWST shared the same standard SED as optically selected AGN \citep{Shen2020}, then they would overproduce the X-ray background by a large factor. However, we have unambiguously shown that the AGN discovered by JWST are much weaker in the X-rays than expected by the standard AGN SED. Regardless of the origin of the X-ray weakness, this indicates that this new population of AGN does not contribute to the X-ray background. The fact that a large population of high-z Compton thick AGN could be accommodated without violating the X-ray background was already pointed out by \cite{Comastri2015}.

\subsection{AGN impact on early galaxy evolution}

The finding of reduced AGN ejective feedback implies that galaxies in the early Universe suffer less of this negative effect on their evolution. This may allow galaxies to grow and form stars faster than predicted by previous models. Additionally, by removing less gas from the circumnuclear region and host galaxy, the lack of strong feedback would allow more fuel to reach the black hole, which could grow more rapidly. This would help to explain the fast growth of black holes in the early Universe.

We finally note that, although the ejective feedback might be reduced, the accreting black holes probably still have significant feedback on their host galaxies in terms of heating, photodissociation, and photoionization of the ISM.

\subsection{Black hole masses}

If the BLR has a covering factor much larger than previous studies, then this would imply that the local virial relations for estimating BH masses, and which use the luminosity of H$\alpha$ as a proxy of the AGN bolometric luminosity (which enters in the equation for estimating the BH masses), should be revisited. In particular, the BH masses should be revised to lower values. However, the BH mass depends on the broad H$\alpha$ luminosity only with a power of 1/2. Hence, if the covering factor of the BLR is higher by a factor between 2 and 10, relative to optical/UV selected AGN (which have been used for the local calibrations), then the BH masses should be revisited downward by a factor between $\sim$1.4 and $\sim$3.

For the same reason, the bolometric luminosities inferred from the broad H$\alpha$
would have been overestimated by a factor between 2 and 10. An implication of this is that the points on the $L_{Bol}/L_X$ vs $L_{Bol}$ diagram (Fig.\ref{fig:Rbol_type1})would move along the diagonal, towards the bottom-left corner, by $\sim$0.3 dex and possibly up to 1 dex. This would bring some of the upper limits consistent with the local relation. However, many objects, and in particular the stack, would still deviate strongly even with the most extreme correction of 1 dex.

Finally, also the inferred accretion rates, in terms of $L/L_{Edd}$, should also be lowered by a similar factor as for the BH masses, i.e. between $\sim$1.4 and $\sim$3.

\subsection{Narrow Line Region and diagnostic diagrams}

If the BLR has a large covering factor, this would imply that only a small fraction of ionizing photons could escape to larger scales to ionize the NLR. If this is the case, then it would provide an additional explanation on why some of the classical narrow line diagnostics fail to identify AGN at high redshift, as a significant fraction of the narrow lines would be dominated by star formation in the host galaxy, despite hosting a powerful AGN.

% \subsection{Circumnuclear little dust medium -> explain BL and MIRI discrepancy in LRD in Pablo's work}

\section{Comparison with previous works }

X-ray weak AGN were obviously found also in the past.
In the case of type 2 AGN, X-ray weakness has often been ascribed to heavy absorption, and generally directly confirmed with hard X-ray spectra, when the signal to noise was adequate \citep[e.g.][]{Salvati1997,Risaliti2000,Maiolino1998,Maiolino2003}. Heavy obscuration in type 2 AGN, is therefore not surprising. What would be surprising, if the Compton thick scenario is confirmed, is the large fraction of Compton thick AGN. As we already mentioned, in the local Universe the fraction of Compton thick AGN is about 50\% \citep{Risaliti1999}, while our findings indicate that the fraction of Compton thick AGN might approach 100\% in high redshift, intermediate/low luminosity AGN. However, other studies have also found a significant evolution of the Compton thick fraction, especially in the low luminosity regime \citep{Ananna2019,Gilli2022}, although not as high as the one inferred by us.
We notice, however, that selecting AGN via the broad H$\alpha$ may bias towards AGN with high EW(H$\alpha$ and, therefore, those with high covering factor of the BLR. Yet, it remains true that, even if biased, this population of AGN discovered by JWST is large in the early Universe, as evinced by the luminosity function presented by \cite{Maiolino_AGN} for the same sample.

The other unexpected result is that type 1 AGN discovered by JWST are extremely X-ray weak too. This is much more unusual in the local Universe, or at high luminosities. Indeed, local type 1 AGN or optically/UV selected quasars at high redshift are characterized by prominent X-ray emission, although becoming gradually dimmer at the highest luminosities, as illustrated in Fig.~\ref{fig:Rbol_type1}.
As mentioned, type 1 BAL quasars are known to be X-ray weak and their weakness is associated with absorption \citep{Gallagher2002,Gallagher2006}, although some BAL quasars appear to be intrinsically weak \citep{Teng2014,LuoBrandt2014}. It is not clear if the type 1 AGN discovered by JWST at high-z are part of the BAL category. Exploring this scenario would require seeing directly the UV emission from the accretion disc, and exploring the presence of blueshifted UV resonant lines (e.g. CIV, SIV, AlIII). However, as already mentioned, most of them, although type 1 (actually they are generally type 1.5--1.9), show some dust reddening in the optical. This implies that the intrinsic UV emission is heavily suppressed and the host galaxy dominates the observed UV light. It is however interesting to note that GN-z11, the highest redshift type 1 AGN, does show CIV blueshifted absorption, similar to BAL quasars. Additionally, the presence of Balmer or HeI absorption clearly indicates that these objects would appear as BAL quasars if observed in the UV without dust obscuration.
Within this context, it is interesting to note that the fraction of (luminous) BAL quasars has been found to increase steeply at high redshift \citep{Bischetti2022,Maiolino2004,Maiolino2001_BAL}.

Additionally, examples of X-ray weak type 1 quasars have been found in the past. \cite{Risaliti2001_Xrayweak}  
explored the properties of emission line (grism) selected quasars and found that they are significantly X-ray weaker relative to the classical quasars selected via their blue colors. They ascribed the difference to dust-poor absorption along the line of sight, which suppresses the X-rays and only mildly redden their optical light.

\cite{PuBrandt2020} also discovered a population of non-BAL X-ray weak quasars, although much smaller ($\sim$5\%) than found among JWST-selected type 1 AGN. They found that a fraction of them tend to be reddened, in line with the absorption scenario. However, they also find that a fraction of them is characterised by weak broad emission lines. This result contrasts with the scenario where the BLR has a large covering factor. However, considering that these UV lines are heavily affected by even small amounts of dust, an absorption scenario may also apply. More recent observations of X-ray weak (type 1) quasars at high energies have revealed the emergence of a very hard component, indicating that their weakness is actually due to a dust poor absorber with large  (Compton thick) column density \citep{WangBrandt2022}.

Interestingly, recent studies have also found that high redshift sources are also significantly affected by X-ray absorption by the inter-galactic medium (IGM) \citep{Arcodia2018,Dalton2022,Gatuzz2024}, and that the absorbing column of gas increases quite steeply with redshift, as $(1+z)^{1.63}$. As the IGM is metal and dust poor, this is another possible source of X-ray weakness, without resulting in significant optical/UV obscuration. However, the IGM absorbing column densities inferred at z$\sim$5 are still of the order a few times $10^{22}~cm^{-2}$, not enough to explain the extreme X-ray weaknesses observed in our sample.

There are also several examples of type 1 X-ray weak AGN and quasars that are likely intrinsically weak, especially in the high accretion rate regime. As already mentioned, NLSy1, which are interpreted as highly accreting AGN, tend to be intrinsically X-ray weak \citep{Vasudevan07}.
\cite{Laurenti2022} analised the X-ray properties of a sample of highly accreting quasars and found that they are intrinsically X-ray weak. X-ray weaknesses in highly accreting AGN is also expected from theoretical modeling \citep{Netzer19,ValianteX2018}.

More recently \cite{PaulBrabndt2024} have identified a population of highly accreting AGN, which are X-ray weak. However, based on their radio properties, they suggest that the X-ray weakness is due to X-ray absorbing gas surrounding the accretion disc.

As already mentioned, the X-ray properties of type 1, JWST selected AGN were already explored by \cite{Yue2024} and \cite{Ananna2024}. However, they limited their analysis to a sample of so-called Little Red Dots, which are only a small fraction ($\sim$10\%--30\%) of the population of JWST-selected AGN at high redshift \citep{Kocevski2024,Greene2024,Hainline2024_LRD}. They suggest intrinsic X-ray weakness as a possible origin. Alternatively, they suggest that the broad component of H$\alpha$ used to identify type 1 AGN might be due to fast galactic outflows (with velocities of a few thousand km/s). However, we have shown that the latter interpretation is untenable, as there is no evidence for such broad component in the forbidden lines (particularly [OIII]), even in the stacked spectra.

Finally, we mention that \cite{Matsui2024} used X-ray stacking of JWST-selected galaxies at 4$<$z$<$7 (expanding on a previous work at lower redshift, based on pre-JWST data, by \citealt{Vito2016}). From their non-detection they set an upper limit on the BH accretion rate density at high redshift. However, if in the intermediate/low luminosity range, high-z AGN are X-ray weak, then the global X-ray properties of high-z galaxies may not be very constraining of the bulk of BH accretion.

\section{Local analogues}
\label{sec:local}

Throughout the paper we have emphasized the ``peculiar'' X-ray weakness of the intermediate/low luminosity AGN discovered by JWST at high-z. One may wonder why this property is found only at high redshift. However, AGN characterized by similar X-ray weakness have actually been found also locally, although possibly overlooked in the past. Leaving aside the large population of heavily absorbed (Compton thick) narrow line, type 2 AGN, there are also a growing number of broad-line, type 1 (or 1.9) AGN that are found to be X-ray weak, as discussed in the following.

We have already discussed the case of NGC~4151, the ``prototypical'' broad-line, Seyfert 1 galaxy, which actually is heavily absorbed in the X-rays \citep{Zoghbi2019_NGC4151}, and also characterized by Balmer absorption \citep{Hutchings2002_NGC4151}, similar to several broad-line JWST AGN at high-z. We have also mentioned that the local population of highly accreting Narrow Line Seyfert 1 are also characterized by steep X-ray spectra, resulting in to X-ray weakness in the hard band that would be probed at high-z. However, there are also other local cases that are resembling the properties of JWST-identified AGN.

NGC~4151 is not the only case of type 1 AGN subject to high absorption in the X-rays. Studying a BAT AGN selected sample \cite{Shimizu2018} found a sample of type 1--1.9 AGN subject to significant X-ray absorption. As in our first intepretation, they suggest that BLR clouds might be responsible for X-ray absorption towards the corona.
A similar finding, i.e. broad-lined AGN with prominent X-ray absorption, was obtained by \cite{Merloni2014} at intermediate redshifts.

Mrk231 is another remarkable local potential analogue, though it is likely representing a different category.  This is the closest (type 1) quasar and it is known to be extremely X-ray weak \citep{Teng2014Mrk231}. Although moderately absorbed (by a partial covering medium), its X-ray weakness does not seem associated with absorption -- it seems an intrinsic property.
\cite{Teng2014Mrk231} suggest that the X-ray weakness might originate from super-Eddington accretion.

Many more cases of type 1 AGN that are anomalously X-ray weak have been found among dwarf, low-metallicity galaxies, which may be more directly connected to the hosts of JWST-identified AGN at high-z. \cite{Simmonds2016Xraydwarfs} studied a sample of metal poor galaxies candidate to host intermediate mass black holes, based on their broad H$\alpha$, and found that they are X-ray undetected, with implied X-ray luminosity 1-2 orders of magnitude below the standard AGN UV-X-ray relations. A similar X-ray weakness was found by \cite{Burke2021Xraydwarf} in a sample of broad-lined H$\alpha$, metal-poor dwarf galaxies. Interestingly, some of the galaxies analysed by \cite{Burke2021Xraydwarf} are characterized by H$\alpha$ absorption, as in JWST-selected AGN at high-z. \cite{Hatano2023SBS} study in detail the very metal poor galaxy SBS 0335 and finding that not only it is characterized by a broad H$\alpha$ \citep[already found in previous studies;][]{Izotov2009SBS}, but also variability in the mid-IR, hence unambiguously confirming the presence of an AGN. However, the nucleus is extremely weak in the X-ray, well below the expectation of a standard AGN SED, and the faint observed X-ray emission can be entirely ascribed to star formation in the host galaxy. Also in the case of SBS 3553 the high resolution spectrum shows indication of Balmer absorption
\citep{{Izotov2009SBS}}, as in some of the high-z JWST-selected AGN.

Finally, \cite{Arcodia2024Xrayweak} explored the X-ray emission of a large sample of dwarf galaxies candidate to host AGN based on their optical/UV variability, and detected only a small minority of them. The stack of the several non-detected is much weaker than expected by the standard disk-corona relations for AGN. They suggest that a canonical X-ray corona might be missing in most of these AGN.

Summarizing, analogues of the X-ray weak AGN found by JWST at high-z are actually present in the local Universe, and they may have been overlooked. The connection between these local and distant populations needs to be explored in greater detail in future works. We however highlight that these potential local analogues provide an excellent (brighter) test bench to explore scenarios proposed for the distant populations.

\section{Conclusions}

We have leveraged the deepest Chandra observations in the GOODS fields to explore 
the X-ray properties of a large sample of 71 type 1 (broad line) and type 2 (narrow line) intermediate-luminosity AGN identified by JWST at 2$<$z$<$11.
These AGN are identified from different parent samples and from different JWST surveys. However, they all have in common
that they have typical luminosities significantly lower than those probed by previous (pre-JWST) surveys, and that they were not pre-selected because known to have an AGN based previous observations at other wavelengths (with the exception of the two X-ray selected objects). Additionally, these JWST-identified AGN are typically hosted in low mass ($M_*\sim 10^8-10^{10}~M_\odot$) and metal poor ($Z\sim 0.1~Z_\odot$) host galaxies, in contrast with more luminous quasars at similar redshifts.

We have obtained the following results:

\begin{itemize}

    \item The vast majority of AGN are undetected. For many of them, the X-ray upper limit is several times, up to two orders of magnitude, below what is expected from the standard SED of classical, optical/UV selected AGN.

    \item Only four objects are detected. Out of these, two were observed spectroscopically because of the previous X-ray detection. The X-ray spectrum of the two which are above the $L_{bol}/L_X$ standard relation reveals that they are Compton thick.

    \item The type 1 AGN with evidence of absorption features in their H$\alpha$ broad profile show some of the most extreme X-ray weaknesses.

    \item The X-ray stacks of the non-detected AGN result in upper limits that are one to two orders of magnitude below what expected from standard AGN SED.

\end{itemize}

We explore the possible scenarios that can be at the origin of the X-ray weakness.

\begin{itemize}

    \item[$\blacksquare$] X-ray absorption

    \begin{itemize}
        \item Absorption by large, Compton thick columns of gas can explain the X-ray weakness.
        \item As the X-ray weakness is also seen in type 1 AGN (which are however typically reddened), this scenario implies absorption by a dust-poor (or dust-free) medium. The BLR clouds are a possible candidate (large column densities and dust-poor/free). However, the large fraction of X-ray weak AGN would imply a large covering factor of the BLR clouds.
        \item The EW(H$\alpha _{broad}$) in JWST-identified AGN is much larger than in standard, optically/UV selected AGN/quasars. This strongly supports the scenario of a BLR with a large covering factor.
        \item The finding of type 1 AGN with evidence of absorption features in their H$\alpha$ broad profile also indicates absorption by gas with densities typical of the BLR.
        \item
        This scenario is further supported by the spectra of the few X-ray detected AGN above the $L_{bol}/L_X$ relation, which indeed reveal Compton thickness.
        \item The [OIII]5007 line profile of JWST-identified AGN does not show the classical blueshifted wing seen low-z AGN and high luminosity quasars, indicating that ionized outflows in JWST-identified AGN are very weak. This suggests much reduced ejective feedback effects in the intermediate AGN luminosity, intermediate BH mass, and low metallicity regime probed by JWST, and suggesting that the dense gas is lingering in the vicinity of the BH, causing the X-ray absorption.
    \end{itemize}

    \item[$\blacksquare$] Intrinsic X-ray weakness

    \begin{itemize}
    
        \item Many of the AGN newly discovered by JWST are Narrow Line Seyfert 1s (i.e. with relatively narrow permitted lines, $FWHM<2000~km/s$, but still broader than the forbidden lines), which are known to have a very steep X-ray spectrum. This can contribute to their X-ray weakness, especially at high-z.
        \item Connected to the previous point, many JWST-identified AGN are accreting at high rates. Since higher BH accretion is expected to result into weaker X-ray emission, this can also contribute to the lack of detections. However, also dormant black holes are found to be very X-ray weak, implying that the accretion rate cannot be the only reason driving X-ray weakness.
        \item Another possibility, that is difficult to test, is that intermediate mass black holes at high redshift have not yet developed properly their corona, possibly because the nuclear magnetic field is not yet strong enough.
        
    \end{itemize}

    \item[$\blacksquare$] Non-AGN sources

    \begin{itemize}
    
         \item Given that the X-ray weakness is seen in the vast majority of JWST-identified AGN at high-z, regardless of their identification method (broad lines, narrow lines, mid-IR emission), it is highly unlikely that all of those selection methods have mis-identifed AGN. The very stringent lower limits on the $L_{Bol}/L_X$ from the stacks would imply that more than 99\% of the sources have been misclassified. Additionally, a number of objects show widespread, multiple, and unambiguous AGN signatures (symmetric broad lines with very high EW, very high ionization lines) and, despite that, are still X-ray undetected.

         \item We have shown that core collapse SNe cannot account for the observed broad H$\alpha$ used to identify many type 1 AGN, based on multiple constraints (lack of significant variability, line luminosity, line profile, absence of other spectral signatures). However, we have shown that a minority of the faintest broad H$\alpha$ with blushifted profiles could be associated with the echoes of superluminous SNe.

         \item The contribution to the broad H$\alpha$ by Very Massive Stars is excluded based on the very high equivalent width of the line.

         \item Tidal disruption events would not be real contaminant, as they are anyway a form of AGN. They are however unlikely to contribute due to the lack of observed variability, the observed line profiles, and because many of them should also be detectable in the X-rays. However, we cannot exclude some minor contribution from this population.

         \item The contribution of galactic outflows is excluded by the lack of any broad component of [OIII]5007, even in the stacked spectra.

        \item We do not exclude that a fraction of the type 2 AGN might actually be extreme star forming galaxies. However, given that their stack results in a very stringent upper limit, it would imply that more than 99\% of them have been misclassified, which is unlikely given that for a good fraction of them the classification is confirmed by more than one diagnostic and/or show high ionization lines. 
         
     \end{itemize}
       
\end{itemize}

Probably, a combination of the scenarios discussed above is jointly responsible for the observed X-ray weaknesses. We stress that better discerning the role of each of these scenarios requires future X-ray observatories more sensitive than Chandra, such as AXIS and Athena \citep{Reynolds2023,Nandra2013,
Marchesi2020}.

Finally, we have also discussed some of the implications of the scenarios that emerged from our results.

 \begin{itemize}

    \item[\ding{70}] As the AGN identified by JWST are observed to be much X-ray weaker than expected by a standard AGN SED, their large number density is not in contrast with the X-ray background constraints.

    \item[\ding{70}] The lack of strong AGN-driven ejective feedback can have important implications on the early evolution of galaxies, allowing them to grow faster and also maintaining more gas available for rapid black hole growth. However, heating and photo-dissociation feedback by the radiation produced by the AGN can still have an impact on the star formation efficiency.

    \item[\ding{70}] If the BLR covering factor is much larger than in the local AGN used to calibrate the BH virial relations, then this implies that the BH masses estimated in the JWST-identified AGN should be revisited downward by a factor of $\sim$1.4--3. For the same reason, the bolometric luminosities inferred from the broad H$\alpha$ should be revisited downward by a factor of $\sim$2--10 and the accretion rates, in terms of $L/L_{Edd}$, should also be revisited downward by a factor of $\sim$1.4--3.

    \item[\ding{70}] Additionally, if the BLR covering factor is very large, this would leave few ionizing photons to escape the nuclear region and produce the Narrow Line Region. This may partly explain why the BPT diagrams of the JWST-selected AGN deviate from the locus populated by AGN locally.

 \end{itemize}

\begin{table*}
\centerline{
\begin{tabular}{lccccccc}
ID  &             z &   log(M$_{\rm BH}$) &  log(L$_{\rm BOL}$) & log(L/L$_{\rm Edd}$) & F(2-10~keV)  &  L(2-10~keV) & L$_{\rm BOL}$/L$_{\rm X}$\\
    \hline
GS 10013704 &      5.919&  7.5&    44.34&  -1.26&  $<$1.40E-17     &$<$3.20E+42 &$>$68  \\
GS 8083 &          4.647&  7.3&    44.55&  -0.80&  $<$5.15E-18     &$<$7.15E+41 &$>$498  \\
GN 1093 &          5.594&  7.4&    44.77&  -0.69&  $<$7.33E-17     &$<$1.50E+43 &$>$39  \\
GN 3608 &          5.269&  6.8&    43.97&  -0.95&  $<$7.33E-17     &$<$1.33E+43 &$>$7.1  \\
GN 11836 &         4.409&  7.1&    44.53&   -0.70& $<$8.97E-18     &$<$1.11E+42 &$>$308  \\
GN 20621 &         4.682&  7.3&    44.66&   -0.74& $<$3.07E-17     &$<$4.33E+42 &$>$105  \\
GN 73488 &         4.133&  7.7&    45.00&   -0.80& $<$8.53E-17     &$<$9.22E+42 &$>$110  \\
GN 77652 &         5.229&  6.9&    44.54&   -0.43& $<$  2.02E-17   &$<$3.61E+42 &$>$95  \\
GN 61888 &         5.874&  7.2&    44.82&   -0.50& $<$3.52E-17     &$<$7.89E+42 &$>$84  \\
GN 62309 &         5.151&  6.6&    44.25&   -0.41& $<$4.01E-17     &$<$6.93E+42 &$>$26  \\
GN 53757 &         4.447&  7.7&    44.44&   -1.35& $<$6.53E-17     &$<$8.27E+42 &$>$34  \\
GN 954 &           6.759&  7.9&    45.62&   -0.38& $<$3.62E-17     &$<$1.02E+43 &$>$409  \\
GS 3073 &          5.55 &  8.2&    45.70&   -1.13&  $<$ 1.64E-17   &$<$3.31E+42 &$>$1514  \\
GN 4014 &          5.228 & 7.58&   44.97&   -0.71& $<$5.84E-17     &$<$1.04E+43 &$>$89  \\
GN 9771 &          5.538 & 8.55&   45.82&   -0.83& $<$3.75E-17     &$<$7.54E+42 &$>$873  \\
GN 12839 &         5.241 & 8.01&   45.49&   -0.62& $<$1.78E-17     &$<$3.19E+42 &$>$978  \\
GN 13733 &         5.236 & 7.49&   44.72&   -0.87& $<$2.60E-17     &$<$4.64E+42 &$>$112  \\
GN 14409 &         5.139 & 7.21&   44.87&   -0.44&  $<$ 2.11E-17   &$<$3.62E+42 &$>$204  \\
GN 15498 &         5.086 & 7.71&   45.02&   -0.79& $<$4.30E-17     &$<$7.23E+42 &$>$144  \\
GN 16813 &         5.355 & 7.55&   44.96&   -0.69& $<$5.58E-17     &$<$1.04E+43 &$>$87  \\
GS 13971 &         5.481 & 7.49&   44.74&   -0.85& $<$2.08E-17     &$<$4.10E+42 &$>$134 \\
GN-z11 &          10.604 & 6.2 &   45.00 &     0.7& $<$1.60E-17     &$<$9.19E+42 &$>$109  \\
%3.00E+43 & $>$33
\hline
ALL           &         5.237   & --  &  45.07&    --  & $<$3.74E-18    &$<$ 3.29E+41 & $>$ 3571\\
ALL-Low-L      &         5.097   & -- &   44.55&    --  & $<$3.43E-18   &$<$ 7.13E+41 & $>$ 497 \\
ALL-High-L     &         5.390   & --  &  45.31&    --  & $<$4.07E-18   &$<$ 4.51E+41 & $>$ 4527\\
\hline
GN 28074 &         2.259   &       8.27&48.23  &-0.75& $<$1.1E-17      & $<2.9$E+41 &$>$15402\\
GN 721   &         2.942   &       7.95&45.82  &-1.21& 1.1$\pm$0.4E-16 & 4.9E+42    &151 \\
GS 49729 &         3.189   &       8.46&42.98  &-1.48& 2.1$\pm$0.3E-15 & 9.9E+43    &13 \\
GS 17341 &         3.598   &       6.52&46.57  &-0.58&$<$1.4E-17       &  $<$1.0e42 &$>$117 \\
GN 2916  &         3.664   &       6.77&45.98  &-1.02&$<$4.6E-17       &  $<$3.5e42 &$>$22 \\
GS 209777 &        3.711   &       8.90&45.42  &-1.59& 1.5$\pm$0.1E-15 &  1.3E+44   & 15 \\
GS 13329 &         3.936   &       6.75&45.30  &-0.72&$<$2.0E-17       &  $<$1.7e42 &$>$85 \\
\hline
XID 403       &         4.76      &       --  &45.30  & --  & 6.6e-16        &  42.83     &$>$ 293\\
GN 1146115    &         6.68          &  8.6 &45.13 & -1.62    & $<$1.5e-17     & $<$4.1e42 & $>$ 329\\
\hline
\end{tabular}}
\caption{Sample properties and results of the X-ray spectral analysis for the type-1 sources. The first group of 21 lines describes the JADES and GA-NIFS sources at redshift $z>4$ In the following three lines we report the values for the stacked spectra relative to the whole sample and the low-luminosity and high-luminosity subsamples. We also include the data for six AGN in the same fields with redshift $2<z<4$, and for two relevant additional sources: XID 403 (\citealt{Gilli2011}, see text for details) and GN 1146115 \citep{Juodzbalis2024}. All upper limits are estimated at a 90\% significance level.}    
\label{tab:type1s}
\end{table*}

\begin{table*}
\centerline{
\begin{tabular}{lccccc}

ID    &   z     &  log(L$_{BOL}$)    &    F(2-10 keV)&  L(2-10 keV) & L$_{BOL}$/L$_X$ \\
\hline
GS 143403   &0.734 & 41.28&$<$ 3.5E-18 &$<$ 7.4E39 &$>$25.7\\
GS 10040620 &1.776 & 41.65&$<$ 5.6E-18 &$<$ 1.3E41 &$>$3.4\\
GS 10012005 &1.862 & 41.46&$<$ 5.9E-18 &$<$ 1.5E41 &$>$1.9\\
GS 8456     &1.884 & 41.58&$<$ 1.5E-17 &$<$ 3.0E42 &$>$0.1\\
GS 209979   &1.896 & 42.70&$<$ 3.1E-17 &$<$ 2.5E42 &$>$2.0\\
GS 10036017 &2.015 & 44.1&$<$ 1.9E-17 &$<$ 6.0E41  &$>$216.5\\
GS 10012511 &2.019 & 41.68&$<$ 7.2E-18 &$<$ 2.2E41 &$>$2.2\\
GS 10008071 &2.227 & 43.14&$<$ 1.6E-17 &$<$ 6.3E41 &$>$21.7\\
GS 8880     &2.327 & 41.97&$<$ 6.3E-18 &$<$ 2.8E41 &$>$3.3\\
GS 17670    &2.349 & 43.01&$<$ 5.9E-18 &$<$ 2.7E41 &$>$38.2\\
GS 10073    &2.631 & 43.00&$<$ 1.0E-17 &$<$ 5.9E41 &$>$16.9\\
GS 10011849 &2.686 & 43.6&$<$ 5.4E-18 &$<$ 3.4E41  &$>$119.7\\
GS 7099     &2.860 & 42.89&$<$ 5.8E-18 &$<$ 4.2E41 &$>$18.5\\
GS 114573   &2.880 & 42.30&$<$ 1.0E-17 &$<$ 4.9E41 &$>$4.1\\
GS 111511   &3.002 & 42.52&$<$ 1.4E-17 &$<$ 7.6E41 &$>$4.4\\
GS 132213   &3.017 & 42.19&$<$ 5.0E-18 &$<$ 2.7E41 &$>$5.7\\
GS 10013597 &3.319 & 42.62&$<$ 3.6E-17 &$<$ 3.8E42 &$>$1.1\\
GS 10035295 &3.588 & 43.72&$<$ 1.4E-17 &$<$ 1.8E42 &$>$29.1\\
GS 104075   &3.719 & 43.16&$<$ 8.0E-18 &$<$ 6.8E41 &$>$21.3\\
GS 108487   &3.974 & 41.93&$<$ 3.8E-17 &$<$ 3.8E42 &$>$0.2\\
GS 7762     &4.145 & 43.64&$<$ 2.2E-17 &$<$ 3.9E42 &$>$11.1\\
GS 95256    &4.162 & 42.10&$<$ 1.0E-17 &$<$ 1.1E42 &$>$1.1\\
GS 8073     &4.392 & 44.4&$<$ 1.1E-17 &$<$ 2.2E42  &$>$130.1\\
GS 10000626 &4.467 & 42.53&$<$ 1.6E-17 &$<$ 3.3E42 &$>$1.0\\
GS 111091   &4.496 & 42.05&$<$ 2.3E-17 &$<$ 3.0E42 &$>$0.4\\
GS 17072    &4.707 & 42.31&$<$ 1.4E-17 &$<$ 3.4E42 &$>$0.6\\
GS 10015338 &5.072 & 43.87&$<$ 4.5E-18 &$<$ 1.3E42 &$>$57.5\\
GS 9452     &5.135 & 43.61&$<$ 1.2E-17 &$<$ 3.5E42 &$>$11.8\\
GS 202208   &5.453 & 44.9&$<$ 3.2E-17 &$<$ 6.2E42  &$>$140.5\\
GS 208643   &5.566 & 43.63&$<$ 1.1E-17 &$<$ 2.3E42 &$>$18.5\\
GS 16745    &5.573 & 43.73&$<$ 1.6E-17 &$<$ 3.6E42 &$>$14.9\\
GS 22251    &5.804 & 44.10&$<$ 8.0E-18 &$<$ 3.1E42 &$>$40.2\\
GS 10056849 &5.820 & 43.18&$<$ 1.9E-17 &$<$ 7.3E42 &$>$2.1\\
GS 201127   &5.837 & 43.47&$<$ 3.0E-17 &$<$ 6.7E42 &$>$4.4\\
GS 99671    &5.936 & 43.19&$<$ 1.8E-17 &$<$ 4.1E42 &$>$3.8\\
GS 9422     &5.942 & 44.49&$<$ 9.3E-18 &$<$ 3.8E42 &$>$81.9\\
GS 10013609 &6.930 & 43.96&$<$ 9.2E-18 &$<$ 5.5E42 &$>$16.5\\
GS 10013905 &7.206 & 43.69&$<$ 5.8E-18 &$<$ 3.8E42 &$>$12.9\\
GS 21842    &7.981 & 43.78&$<$ 4.9E-18 &$<$ 4.0E42 &$>$15.1\\
GS 10058975 &9.436 & 44.38&$<$ 8.9E-18 &$<$ 3.0E43 &$>$8.1\\
\hline
STACK1   &(z$<$3, @z=2) & 42.55 &$<$ 6.3E-19 &$<$ 1.3E40 &$>$ 273\\
STACK2  & (z$>$3, @z=5) & 43.60 &$<$ 1.4E-19 &$<$ 2.0E40 &$>$ 990\\
\hline
GN 42437    &5.59  & 44.40&$<$ 7.4E-17  & $<$1.4E43 & $>$17.9\\
GS 21150         &3.8   & 43.80&1.6E-16  & 2.0E42 & 31.5\\
\hline
\end{tabular}}
\caption{Sample properties and results of the X-ray spectral analysis of the sample of type 2 AGNs. $^a$: Flux and luminosity obtained from a fit with a power law with slope $\Gamma=1.7$. All upper limits are estimated at a 90\% significance level.
%$^b$: intrinsic flux and luminosity obtained from a fit with a reflection model (BORUS, Balokovic et al. 2019). \RMcomm{There should be an X-ray detected source in the Scholtz+ type 2 sample}
}    
\label{tab:type2s}
\end{table*}

\section*{Acknowledgements}

RM,IJ,JS,H\"U,FD and acknowledge support by the Science and Technology Facilities Council (STFC), by the ERC through Advanced Grant 695671 ”QUENCH”, and by the UKRI Frontier Research grant RISEandFALL. RM also acknowledges funding from a research professorship from the Royal Society.
EB acknowledges financial support from INAF under the Large Grant 2022 ``The metal circle: a new sharp view of the baryon cycle up to Cosmic Dawn with the latest generation IFU facilities''. 
H{\"U} gratefully acknowledges support by the Isaac Newton Trust and by the Kavli Foundation through a Newton-Kavli Junior Fellowship.
MS acknowledges financial support from the Italian Ministry for University and Research, through the grant PNRR-M4C2-I1.1-PRIN 2022-PE9-SEAWIND: Super-Eddington Accretion: Wind, INflow and Disk-F53D23001250006-NextGenerationEU.

%%%%%%%%%%%%%%%%%%%%%%%%%%%%%%%%%%%%%%%%%%%%%%%%%%
\section*{Data Availability}

The X-ray datasets were obtained from the Chandra X-ray Centre
https://cxc.harvard.edu/cda/
and JWST spectroscopic data are publicly available at MAST STScI archive https://archive.stsci.edu/missions-and-data/jwst.



\bibliographystyle{mnras}
\bibliography{example,roberto_GNz11,ignas,xray} % if your bibtex file is called example.bib

%\begin{thebibliography}{99}
%\end{thebibliography}

%%%%%%%%%%%%%%%%%%%%%%%%%%%%%%%%%%%%%%%%%%%%%%%%%%

%%%%%%%%%%%%%%%%% APPENDICES %%%%%%%%%%%%%%%%%%%%%

\appendix

\section{X-ray spectra of detections and stacks}

\begin{figure}
	\includegraphics[width=0.9\linewidth]{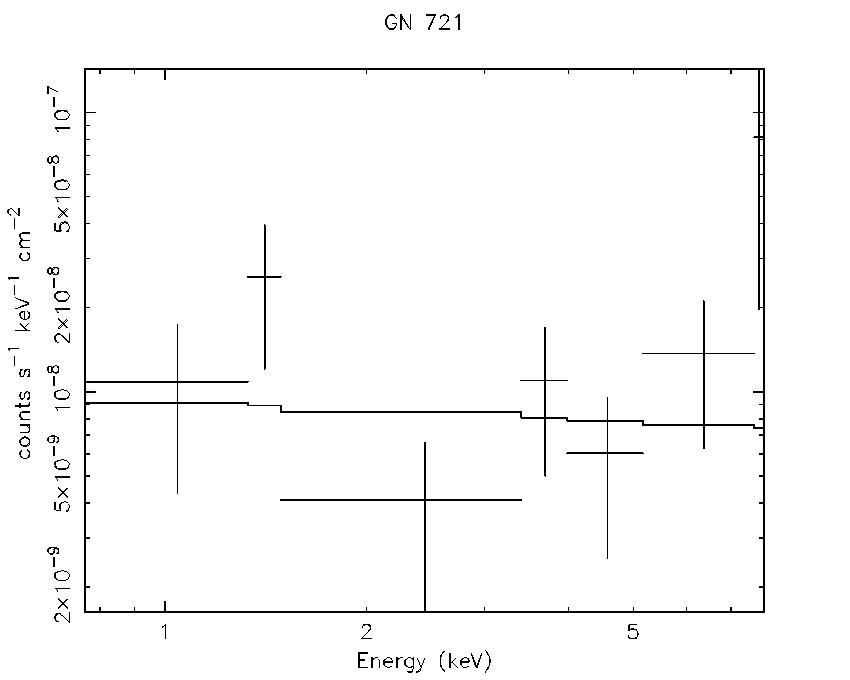}
\includegraphics[width=0.9\linewidth]{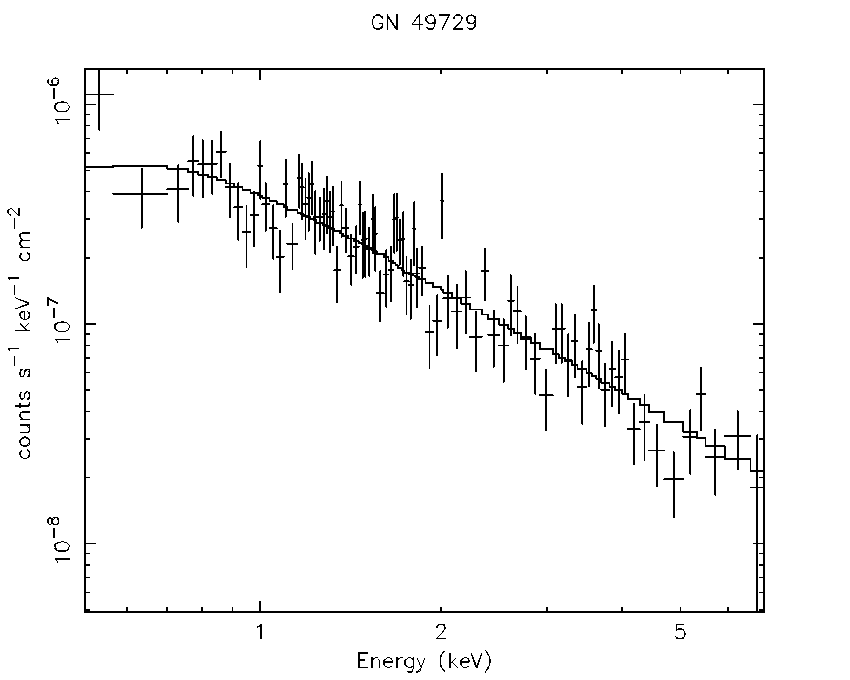}
\includegraphics[width=0.9\linewidth]{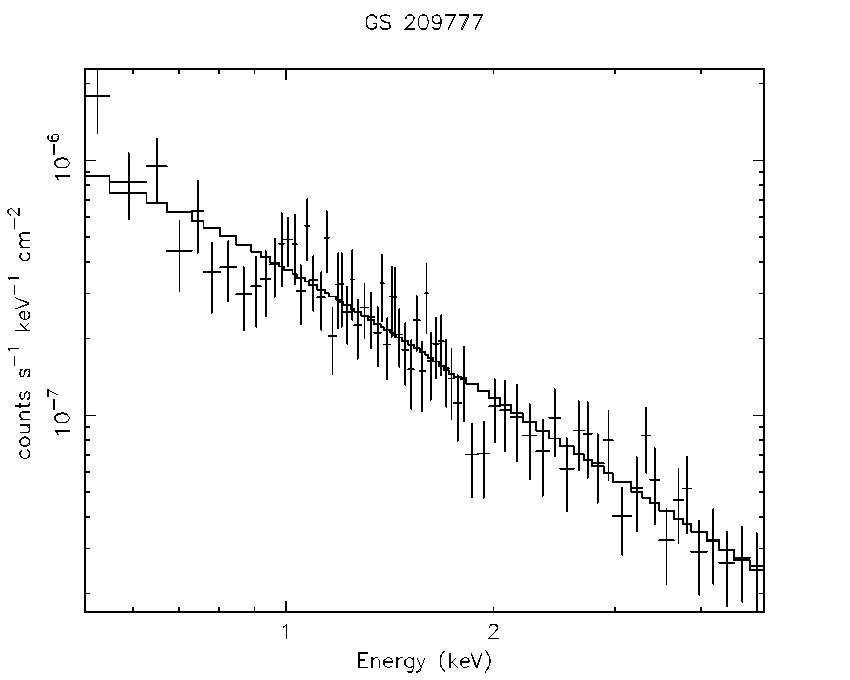}
	\caption{Spectra of the three detected objects: GN~721 (top panel), GS~49729 (middle panel) and GN~209777 (bottom panel). The low signal-to-noise detection of GN~721 only allows a rough estimate of the photon index $\Gamma=0.1\pm0.6$, suggesting a reflection-dominated spectrum (hence a spectrum in which the direct component is likely completely absorbed by a Compton thick medium). The spectrum of GS~49729 is reproduced by a power law with $\Gamma=1.6\pm0.05$ and absorbing column density $N_H<4\times10^{22}$erg cm$^{-1}$s$^{-2}$. Similarly, the spectrum of GS~209777 is reproduced by a power law with $\Gamma=1.7\pm0.1$ and absorbing column density $N_H<3\times10^{22}$erg cm$^{-1}$s$^{-2}$.}
	\label{fig: Xspectra_det}
\end{figure}

\begin{figure}
	\includegraphics[width=1.0\linewidth]{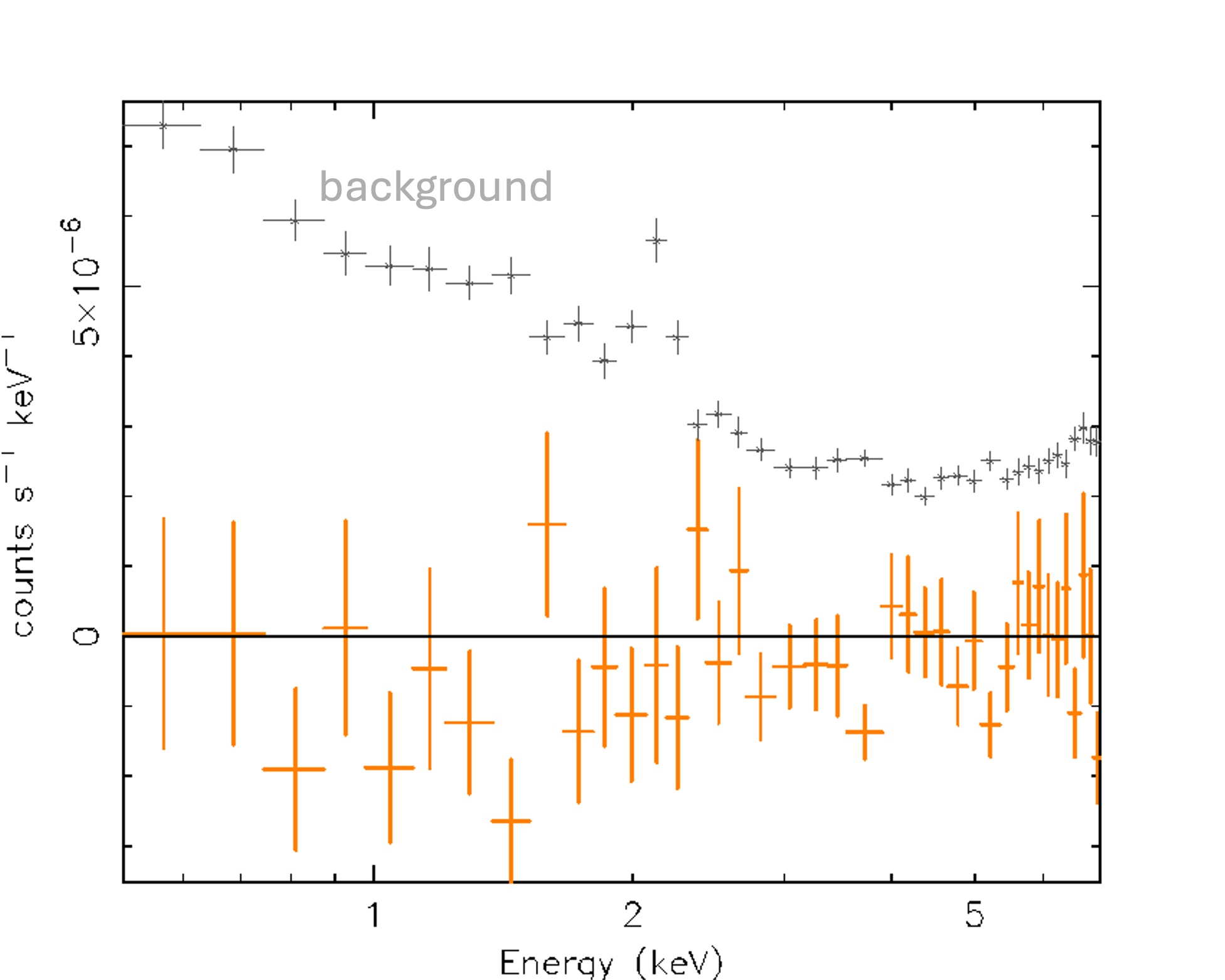}
\includegraphics[width=1.0\linewidth]{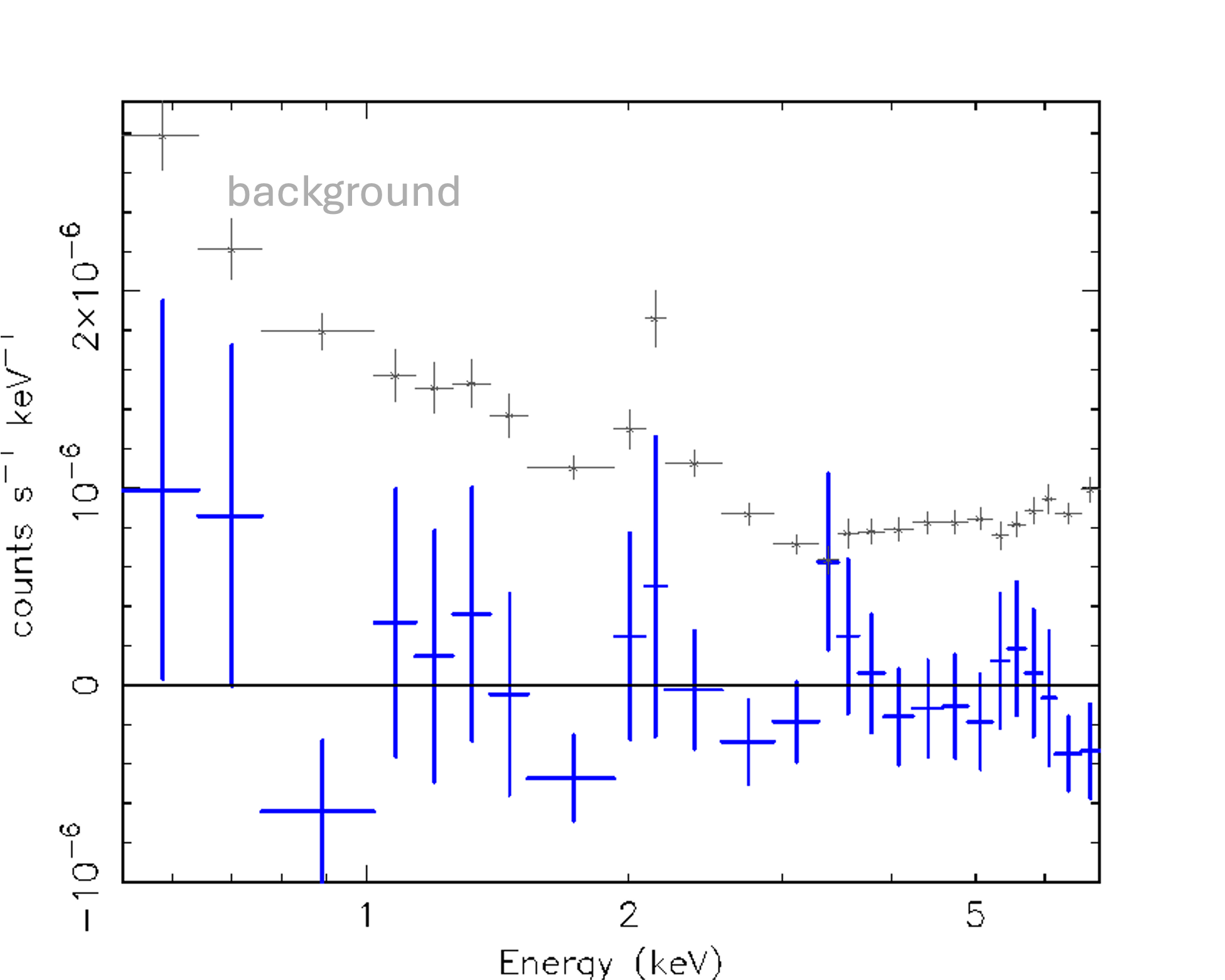}
	\caption{Stacked X-ray spectra for the high-luminosity (top panel) and low-luminosity (bottom panel) type 1 subsamples at z$>$4. The grey points show the background level. In both cases the stacked spectra are consistent with zero net counts, and are well below the background level.} 
	\label{fig: Xspectra_bins}
\end{figure}

In this Appendix we show the spectra of the two detected sources (Fig.~\ref{fig: Xspectra_det}) and of the stacked spectra of the two luminosity bin $\log(L_{BOL})<44.9$ and $\log(L_{BOL})>44.9$ for the type 1 AGN at z$>$4 (Fig.~\ref{fig: Xspectra_bins}). 

\label{app:stack_bins}

\section{\texorpdfstring{EW(H$\alpha$)}{EW(Ha)} versus accretion rate}
\label{app:ew_acc_rate}

\begin{figure}

	\includegraphics[width=1.0\linewidth]{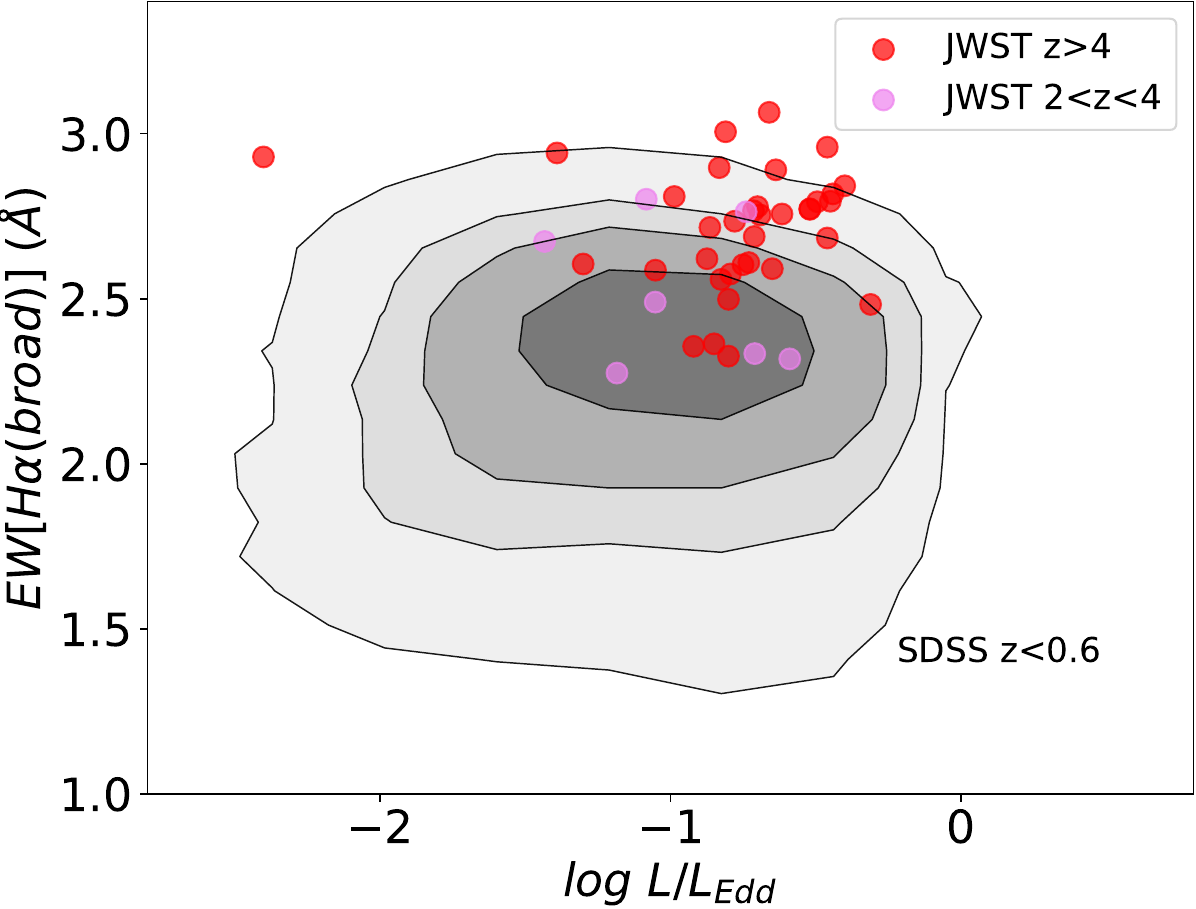}
	\caption{Equivalent width of the broad component of H$\alpha$ versus AGN accretion rate, $L/L_{Edd}$, for the AGN discovered by JWST at high redshift (red for those at z$>$4, violet for those at 2$<$z$<$4), compared with AGN and quasars at low redshift (gray contours).} 
	\label{fig:ew_LLEdd}
\end{figure}

Fig.~\ref{fig:ew_LLEdd} shows the distribution of $EW(H\alpha _{broad})$ as a function of accretion rate, $L/L_{Edd}$. The gray-shaded contours show the distribution of AGN and quasars at low redshift, from \cite{Lusso2020}. The circles show the distribution of JWST-discovered AGN, red for those at z$>$4 and violet for those at 2$<$z$<$4.. The local distribution does not show any significant correlation, as already pointed out by \cite{Ferland2020}. The JWST-identified AGN are, on average, located above the local distribution, regardless of 
$L/L_{Edd}$.

\section{Spectra stacking methodology}
\label{app:stack}

In this appendix we provide additional information on the stacking procedure.
All spectra were shifted to rest frame and rebinned to a common wavelength grid, designed so as to preserve the observed frame resolution also in the rest frame. The spectra are then normalized by their flux at a reference wavelength (generally 5100$\AA$) in order not to bias the stack towards the most luminous objects. In each spectral channel a bootstrap procedure was adopted to evaluate both the average flux and the uncertainty. In brief, a target number of average spectra is decided before the stacking (generally $N_{target}$ = 100). Then, $N_{draw}$ spectra, with $N_{draw}$ equal to the total number of spectra in each sample, are randomly extracted allowing for replacement and stacked together taking the median flux in each spectral channel (less prone to extreme values) to create the $i-th$ composite spectrum.  This procedure was performed $N_{target}$ times, building the distribution of the mean fluxes in each spectral channel. The final spectrum in each spectral channel was obtained as the mean of the mean fluxes and the uncertainty as the standard deviation.

% \section{Calculation of the SN constraints}
% \label{app:sn_comp}

%%%%%%%%%%%%%%%%%%%%%%%%%%%%%%%%%%%%%%%%%%%%%%%%%%

% Don't change these lines
\bsp	% typesetting comment
\label{lastpage}
\end{document}